%% file: main.tex
\documentclass{article}

\usepackage{arxiv}

\usepackage[utf8]{inputenc} 
\usepackage[T1]{fontenc}    
\usepackage{hyperref}       
\usepackage{url}            
\usepackage{booktabs}       
\usepackage{amsfonts}       
\usepackage{nicefrac}       
\usepackage{microtype}      
\usepackage{lipsum}		
\usepackage{graphicx}
\usepackage{doi}
\usepackage{multirow}
\usepackage{amsmath, amsthm, amssymb}
\usepackage[sorting=none, style=nature]{biblatex}
\title{GAN-based Tabular Data Generator for Constructing Synopsis in Approximate Query Processing: Challenges and Solutions}

\author{ \href{https://orcid.org/0000-0002-0506-9758}{\includegraphics[scale=0.06]{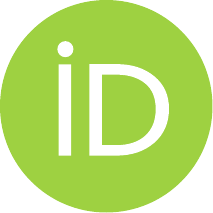}\hspace{1mm}Mohammadali Fallahian}\thanks{Corresponding author.} \\
	Department of Computer Science\\
	University of North Carolina at Charlotte\\
	Charlotte, NC, USA \\
	\texttt{mfallahi@uncc.edu} \\
	\And
    Mohsen Dorodchi\\
	Department of Computer Science\\
	University of North Carolina at Charlotte\\
	Charlotte, NC, USA \\
	\texttt{mdorodch@uncc.edu} \\
    \And
    Kyle Kreth\\
	Department of Computer Science\\
	University of North Carolina at Charlotte\\
	Charlotte, NC, USA \\
	\texttt{kekreth@uncc.edu} \\
}

\renewcommand{\shorttitle}{GAN-based Tabular Data Generator for Constructing Synopsis}

\hypersetup{
pdftitle={GAN-based Tabular Data Generator for Constructing Synopsis in Approximate Query Processing: Challenges and Solutions},
pdfsubject={GANs, APQs},
pdfauthor={Mohammadali Fallahian},
pdfkeywords={Generative Adversarial Networks (GANs), Approximate Query Processing (AQP), Data synopsis, Tabular generator},
}

\begin{document}
\maketitle

\begin{abstract}
	In data-driven systems, data exploration is imperative for making real-time decisions. However, big data is stored in massive databases that are difficult to retrieve. Approximate Query Processing (AQP) is a technique for providing approximate answers to aggregate queries based on a summary of the data (synopsis) that closely replicates the behavior of the actual data, which can be useful where an approximate answer to the queries would be acceptable in a fraction of the real execution time. This study explores the novel utilization of Generative Adversarial Networks (GANs) in the generation of tabular data that can be employed in AQP for synopsis construction. We thoroughly investigate the unique challenges posed by the synopsis construction process, including maintaining data distribution characteristics, handling bounded continuous and categorical data, and preserving semantic relationships and then introduce the advancement of tabular GAN architectures that overcome these challenges. Furthermore, we propose and validate a suite of statistical metrics tailored for assessing the reliability of the GAN-generated synopses. Our findings demonstrate that advanced GAN variations exhibit a promising capacity to generate high-fidelity synopses, potentially transforming the efficiency and effectiveness of AQP in data-driven systems.
\end{abstract}


\section{Introduction}
\input{UpdatedSection/1-Introduction}

\section{Background}
\input{UpdatedSection/2-Background}

\section{Synopsis Construction Challenges}
\input{UpdatedSection/3-SynopsisChallenges}

\section{GAN-based Synopsis Construction Solutions}
\input{UpdatedSection/4-Solution}

\section{Synopsis Evaluation and Error Estimation}
\input{sections/9-Evaluation}

\section{Conclusion}
\input{sections/10-Conclusion}

\printbibliography

\end{document}

%% file: UpdatedSection/1-Introduction.tex
Research and business today rely heavily on big data and its analysis. However, big data is stored in massive databases that are difficult to retrieve, analyze, share, and visualize using standard database query tools \cite{sagiroglu2013big}. For data-driven systems, data exploration is imperative for making real-time decisions and understanding the knowledge contained in the data. However, supporting these systems can be costly, especially regarding big data. One of the most critical challenges posed by big data is the high computational cost associated with data exploration and real-time query processing \cite{li2018approximate}. To assist with the analysis of big data, several systems have been developed, such as Apache Hive, which typically takes a considerable amount of time to respond to analytical queries \cite{muniswamaiah2020approximate}. However, approximation results can sometimes be provided for a query in a fraction of the execution time to resolve this issue, particularly in aggregation queries. This is because aggregation queries are typically designed to provide a big picture of a large amount of information without having to compute an exact answer \cite{hellerstein1997online}. The majority of analytical queries require aggregate answers (such as sum(), avg(), count(), min(), and max()) for a given set of queries (joined or nested queries) over one or more categories (Group by columns) on a subset (where and having) of big data. Approximate Query Processing (AQP) comes to the rescue by identifying a summary of the population (a.k.a synopsis) for discovering trends and aggregate functions results \cite{chaudhuri2017approximate}. Online aggregation and Offline precomputed synopsis are the two primary categories that can be used to classify existing AQP approaches. Offline techniques summarize the data distribution and return the approximate results by running queries on these synopses. However, Online aggregation techniques progressively generate synopses and return approximate results while data is processing. The traditional approach in both categories uses data distribution to generate a subset of data with statistical methods such as sampling methods \cite{li2018approximate}. One novel technique for AQP is to take advantage of machine learning to further reduce the execution time, improve accuracy, and support all types of aggregate functions. For instance, the DBEst Query processing engine \cite{ma2019dbest} trains models, notably regression models and density estimators, that provide accurate, efficient, and cost-effective responses to different types of aggregate queries. Learning-based AQP (LAQP) \cite{zhang2021laqp} and ML-AQP \cite{savva2020ml} methods build machine learning models based on historical executed queries. The former builds an error model to predict each incoming query's sampling-based estimation error, whereas the latter trains models that learn patterns to predict future query results with a bound error by applying prediction intervals constructed using Quantile Regression models. Deep Generative Models (DGM) are employed to approximate complex, high-dimensional probability distributions of the population. Therefore, DGMs can be used to estimate each observation's probability and generate data synopses from the underlying distribution \cite{ruthotto2021introduction}. Thirumuruganathan et al. \cite{thirumuruganathan2020approximate} utilized the Deep Generative Model in the AQP with Variational Autoencoders (VAEs) to generate many samples as required without accessing the underlying dataset. VAEs are a type of generative model in which auto-encoders produce new data from an interpretable latent space by encoding the input distribution. The Generative Adversarial Network (GAN) is another state-of-the-art algorithm that follows a direct implicit density model that samples directly from the model provided distribution \cite{goodfellow2016nips} without estimating the data distribution \cite{gui2021review}. Consequently, GANs appear to be a suitable option for AQP.\\
The remainder of this paper is structured to methodically explore the intersection of synopsis creation in APQ and Generative Adversarial Networks (GANs). Section 2 explores the theoretical underpinnings of database synopses and AQP, alongside a technical exposition on tabular GANs, setting the stage for understanding their relevance and application. Section 3 identifies and discusses the inherent challenges in constructing synopses from relational databases, underscoring the need for innovative approaches. Section 4 proposes a GAN-based solution, demonstrating how tabular GAN-based generators can effectively meet these synopsis creation challenges. Section 5 details the evaluative metrics and methodologies for assessing the fidelity and utility of the generated synopses, including error estimation techniques. The paper concludes with Section 6, which synthesizes our findings and reflections on the potential of tabular GANs to enhance real-time decision-making in data-intensive environments.

%% file: UpdatedSection/2-Background.tex
This section provides the necessary foundation for comprehending the fundamental principles underlying the data synopses in APQ, and the novel utilization of GANs in the context of tabular data.
\subsection{Data Synopsis in Database}
Query processing refers to the process of the compilation and execution of a database query using a specific query language, such as SQL, in order to obtain an approximate result of the requested query. Initially, the query parser validates the query to ensure that the query has been properly stated. Afterward, the query optimizer adjusts the plan to provide a more effective query execution plan. Finally, the query evaluation and execution engine executes the query on the database and returns the results \cite{Markl2009}. A traditional database system performs aggregate operations in batch mode, in which a query is submitted, the system processes a huge amount of data slowly and then returns the final result \cite{hellerstein1997online}. Result in, the primary concern of query processing is how to process queries efficiently based on computational resources and time. Occasionally, it is impossible to provide exact results in a reasonable amount of time, and an approximate answer with some error guarantee would greatly assist users. In order to approximate a query plan outcome for complex join queries, the optimizer requires accurate estimates of the sizes of results generated at accurate selectivity estimates. As a result, data synopses can be used to estimate the number of results generated by a query by estimating the underlying data distribution \cite{spiegel2006graph}.
\subsection{Approximate Query Processing (AQP)}
\begin{figure}[ht]
    \fbox{\includegraphics[width=\dimexpr\textwidth-2\fboxrule-2\fboxsep]{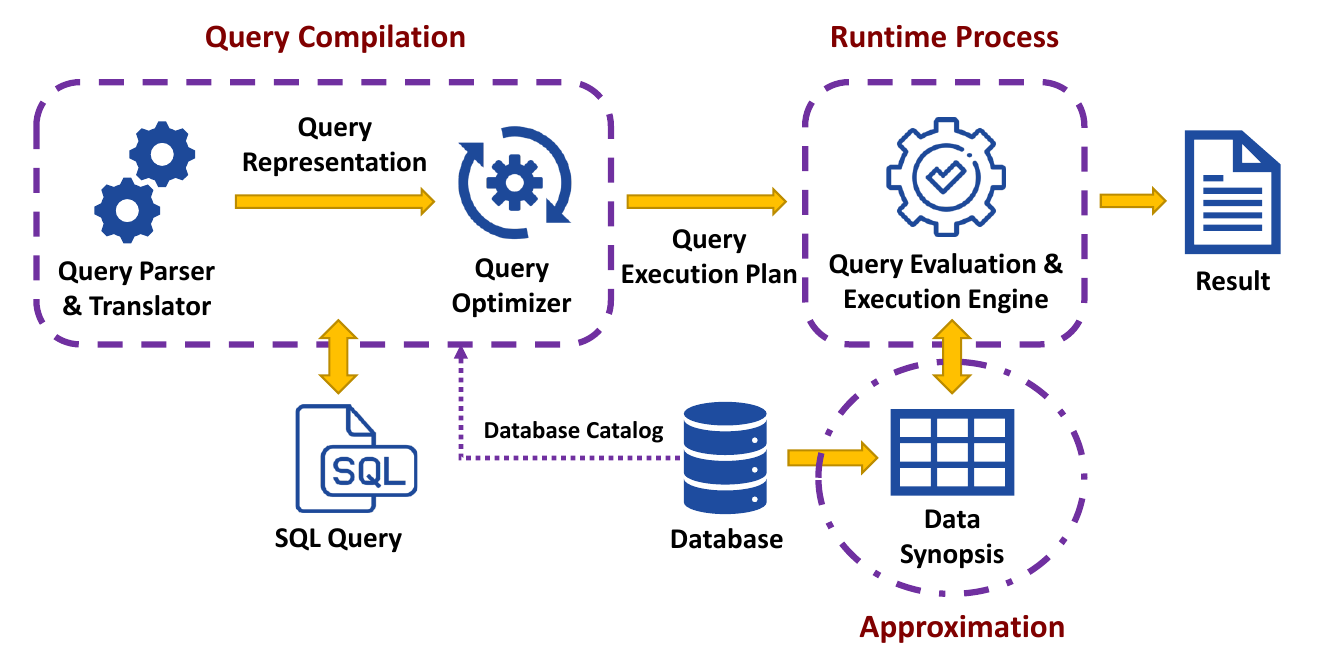}}
    \centering
    \caption{Query processing flow diagram in APQ.}
    \label{fig:APQ}
\end{figure}

Approximate Query Processing (AQP) is a method that returns approximations of aggregate query answers using a data synopsis that closely replicates the actual data's behavior \cite{liu2009approximate}. As a higher level of abstraction, AQP aims to calculate an answer that is approximate to the actual query result based on a data synopsis as a highly compressed and lossy version of the database \cite{spiegel2009tug}. In Figure \ref{fig:APQ}, the different phases of query processing are shown, as the query in AQP is executed based on a data synopsis rather than actual data.
Based on a cost-effective approach, approximation accuracy (consequently completion time) is determined by the size of data synopses, which means how much smaller the synopsis is than the original database \cite{liu2009approximate}. 
We can create this synopsis using either offline or online techniques. Offline synopsis is built using existing data statistics and helps answer queries quickly, but can involve more complex and resource-intensive methods. With offline methods, database optimization techniques like replication and indexing can be employed to refine the synopsis when the database changes \cite{mozafari2015handbook}. On the other hand, online synopsis allows for real-time query monitoring, giving users preliminary results that are refined as more data is processed, and stopping once the results reach a satisfactory level of accuracy and confidence \cite{hellerstein1997online}.\\
By taking an online approach, there is no need to make any a priori assumptions. In contrast to the offline approach, creating good data synopses is much more difficult \cite{mozafari2015handbook}. The Online Analytical Processing (OLAP) system is an example of these systems, and one of its key issues is the regular updating of aggregates to ensure that approximated answers are smooth and continuously improving. By constructing a concise and accurate synopsis of the underlying data distribution, the system consistently strives to reduce the amount of time it takes to complete the task \cite{li2018approximate}.
\subsection{Synopsis Construction}
There may be considerable differences in the structure of the synopsis, and it should be tailored to the problem being addressed. As an example, the AQP synopsis structure is likely to differ from data mining tasks such as change detection and classification \cite{aggarwal2007survey}. AQP systems should generate an effective synopsis that can be applied to various data distributions and data types within different databases. It is common for big data to produce massive amounts of complex data in a streaming manner. Traditionally, streaming algorithms are evaluated based on three factors: running time, memory complexity, and approximation ratio \cite{tan2022one}. Synopsis construction in data streams can be achieved using a variety of techniques:\\
\textbf{Sampling methods:} It has been demonstrated that sampling is a simple and effective method of obtaining approximate results providing an error guarantee when compared with other approximate query processing techniques. It is possible to divide a sampling estimation roughly into two stages. Initially, a suitable sampling method must be identified to construct a sampling synopsis from the original data set, and then a sampling estimator must be analyzed in order to determine its distribution characteristics \cite{Zhang2009}.\\
\textbf{Histograms:} In the histogram, the value range of attributes is divided into K buckets with equal widths, and then the numbers of values falling within each bucket are counted \cite{piatetsky1984accurate}. Based on these statistics, the histogram can then be used to reconstruct the value of the entire dataset within each bucket using the most representative statistics for each bucket \cite{li2018approximate}. In real-world applications, multiple visits to a data stream can improve accuracy and performance, but this is not realistic. For this reason, one-pass and high-accuracy algorithms are required in order to generate data synopses \cite{Zhang2009}. Histogram is cheap to compute since only one pass through the relationship is required, but its precision is not always satisfactory \cite{piatetsky1984accurate}.\\
\textbf{Wavelets:} In synopsis construction, wavelets are derived from wavelet transformations in signal processing, which decompose a function into a set of wavelets using a wavelet decomposition tree. In order to generate a synopsis of data, the original data is decomposed n times using the approximation coefficient at each level of the tree to reach an abstraction of data \cite{RUSSELL20082316}. Wavelets are conceptually similar to histograms, but a wavelet transforms data and attempts to compress the most expressive features in data, requiring more computation time, as opposed to a histogram, which generates buckets by analyzing a subset of the original data in a short amount of time \cite{li2018approximate}.\\
\textbf{Sketches:} Sketches are a type of probabilistic data structure based on the frequencies of unique items in a dataset \cite{yang2017sf}. In order to construct the synopses, k random vectors can be selected and the data can be transformed by dot product to those vectors \cite{aggarwal2007survey}.\\
Although this section introduced the basic methods for constructing synopses, many other techniques, such as clustering \cite{aggarwal2007survey} and materialized views \cite{halevy2001answering}, can also be used to generate them. Traditional methods have many challenges relating to data type, structure, distribution, and query aggregation functions. Furthermore, synopses provide the most accurate summary using the entire data stream, and it would be inconvenient to retrieve the entire data set in real-time databases as it changes over time. A discussion of the challenges associated with generating data synopses in relational databases will be presented in the following subsection.
\subsection{GAN-based Tabular Generator}
\begin{figure}[ht]
    \fbox{\includegraphics[width=\dimexpr\textwidth-2\fboxrule-2\fboxsep]{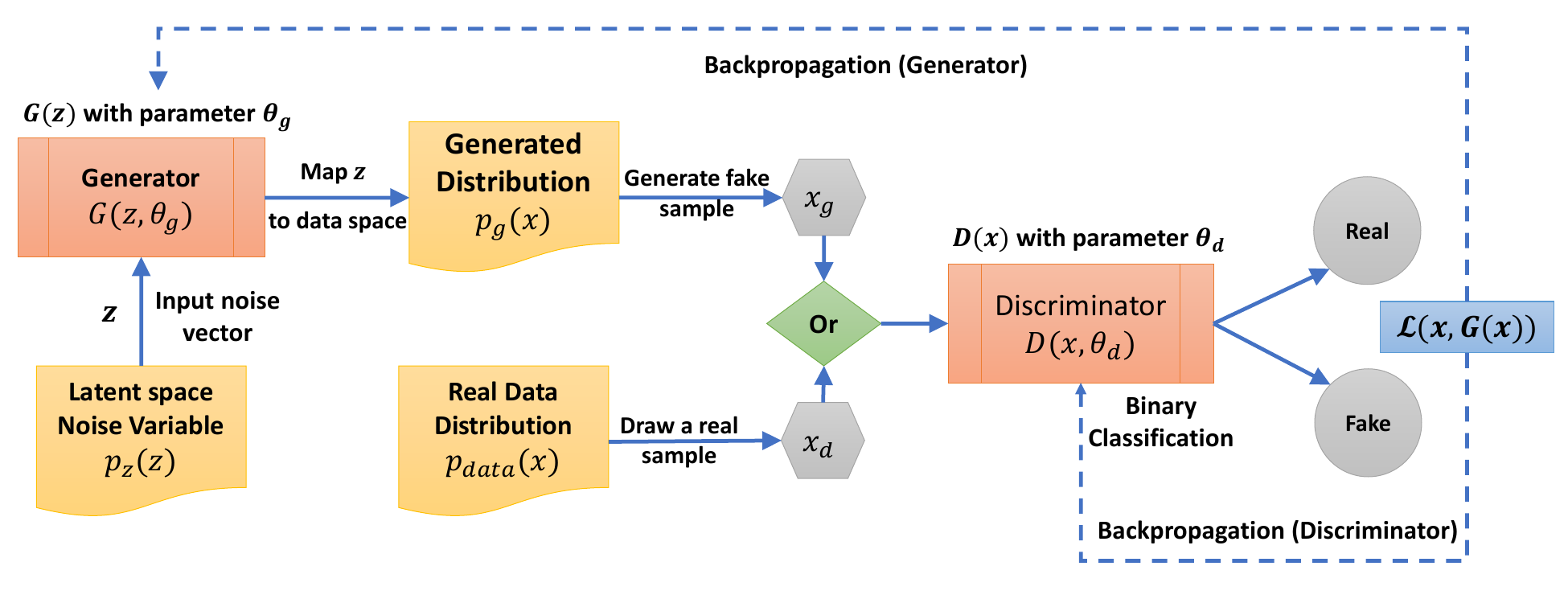}}
    \centering
    \caption{GANs process flow diagram.}
    \label{fig:gan-architecture}
\end{figure}

\begin{figure}[ht]
    \fbox{\includegraphics[width=\dimexpr\textwidth-2\fboxrule-2\fboxsep]{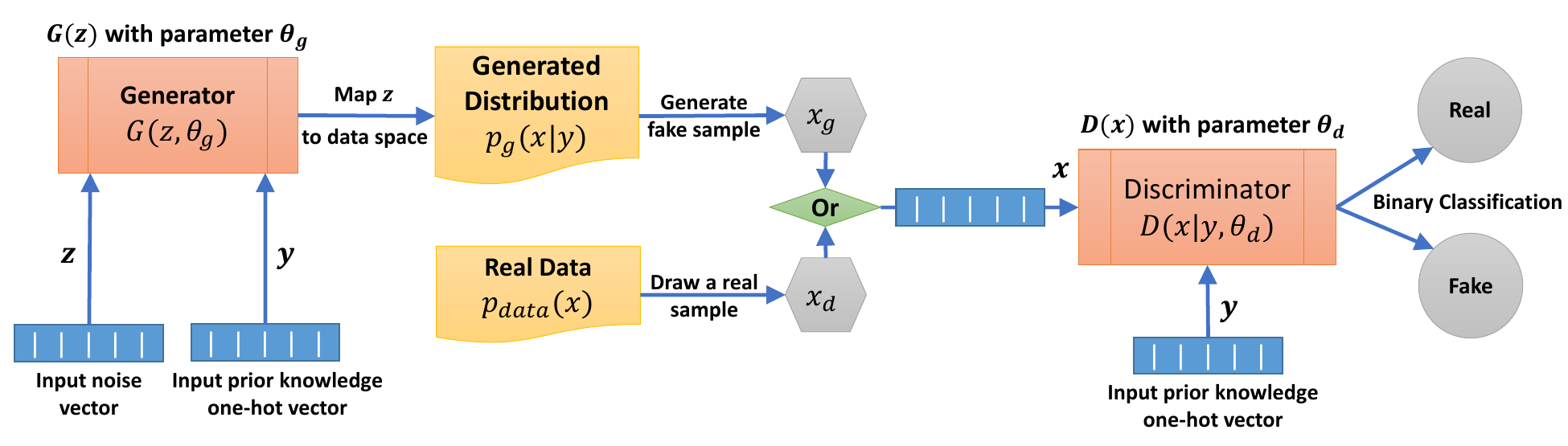}}
    \centering
    \caption{Conditional GAN process flow diagram.}
    \label{fig:cgan-architecture}
\end{figure}

\begin{figure}[ht]
    \fbox{\includegraphics[width=\dimexpr\textwidth-2\fboxrule-2\fboxsep]{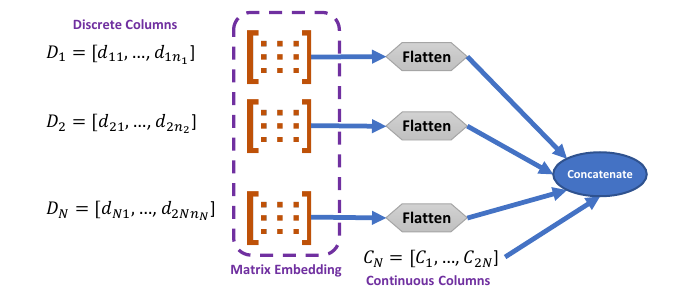}}
    \centering
    \caption{Pre-processing input data before feeding the discriminator in PNR-GAN}
    \label{fig:MatrixEmbedding}
\end{figure}

\begin{figure}[ht]
    \fbox{\includegraphics[width=\dimexpr\textwidth-2\fboxrule-2\fboxsep]{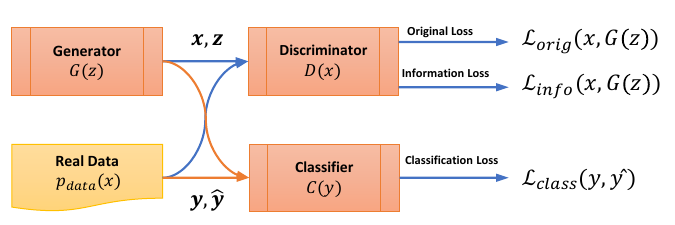}}
    \centering
    \caption{Loss functions representation in table-GAN architecture.}
    \label{fig:tableGAN}
\end{figure}

GANs were introduced in computer vision, where they are commonly used to process image data via Convolutional Neural Networks (CNNs). However, they are capable of generating tabular data as well. The GAN architecture has undergone numerous enhancements in recent years as a result of the improvement in the architecture among the research community over the past few years \cite{wang2021generative}. To determine whether or not GAN is an appropriate option for synopsis generation, first we provide a detailed description of the GAN method and its architecture.\\
Generative Adversarial Networks (GANs) are characterized by two neural networks, the generator, which creates data that is intended to mimic the true data distribution, and the discriminator, which evaluates the data, distinguishing between the generator's fake data and real data from the actual distribution \cite{goodfellow2014generative}. The generator draws a random vector $z$ from the latent space with the distribution $p_z(z)$. The generator $G(z;\theta_g)$ then uses a parameter $\theta_g$ to map $z$ from the latent space to the data space. Therefore, $p_g(x)$ the probability density function over the generated data is used by $G(z)$ to generate $x_g$. Then, the discriminator neural network $D(x;\theta_d)$ receives randomly either $x_g$ the generated sample or $x_{data}$ the actual sample from the probability density function over the data space $p_{data}(x)$. The discriminator neural network $D(x;\theta_d)$ is a binary classification model in which $D(x)$ returns the probability that $x$ is derived from real data. Therefore, the output of this function is a single scalar that indicates if the passed sample is real or fake. Figure \ref{fig:gan-architecture} depicts the described process and GANs architecture. $\theta_g$ and $\theta_d$ are the weights for the generator and discriminator that are learned through the optimization procedure during training.
The goal of the discriminator in training is to maximize the probability that a given training example or generated sample has been assigned the proper label, whereas the goal of the generator is to minimize the probability that it has detected real data. Therefore, the objective function can be expressed as a minimax value function, $V(G,D)$, which is jointly dependent on the generator and the discriminator, where:   
\begin{equation}\label{eq1}
\begin{aligned}[b]
& \underset{G}{\mathrm{min}}\ \underset{D}{\mathrm{max}}V(D,G)=\mathbb{E}_{x\sim p_{data}(x)}[\log(D(x))]+\mathbb{E}_{z\sim p_z(z)}[\log(1-D(G(z)))]
\end{aligned}
\end{equation}
The discriminator performs binary classification, which gives a value of 1 to real samples ($x\sim p_{data}(x)$) and a value of 0 to generated samples ($z\sim p_z(z)$). Therefore, in the optimal adversarial networks, $p_g$ converges to $p_{data}$, and the algorithm is stopped at $D(x)=1/2$ which means the global optimum occurs when $p_g=p_{data}$ \cite{goodfellow2014generative}.\\
The generating data in an unconditioned GAN is completely unmanageable in multimodal distribution. Mirza et al. \cite{mirza2014conditional} introduced a conditional version of GAN that can provide generators with prior information so that they can control the generation process for different modes. Achieving this objective requires conditioning the generator and discriminator on some additional information, $y$, where $y$ can be anything from class labels to information on the distribution of data (modes). This can be done by giving the discriminator and the generator $Y$ as an extra input layer in the form of a one-hot vector. In fact, the input noise $p_z(z)$ to the generator is not truly random if the information $y$ is added to it, and the discriminator does not only regulate the similarity between real and generated data, but also the correlation between the generated data and input information $y$.
Therefore, the objective function in Eq. \ref{eq1} can be rewritten as follows:   
\begin{equation}\label{eq-CGAN}
\begin{aligned}[b]
& \underset{G}{\mathrm{min}}\ \underset{D}{\mathrm{max}}V(D,G)=\mathbb{E}_{x\sim p_{data}(x)}[\log(D(x|y))]+\mathbb{E}_{z\sim p_z(z)}[\log(1-D(G(z|y)))]
\end{aligned}
\end{equation}
Figure \ref{fig:cgan-architecture} illustrates the structure of a CGAN and how to input information is applied during the process.
A majority of applications for conditional GAN were concerned with synthesizing images by giving the label for the image that should be generated. Nonetheless, in the case of tabular data, this could be the shape of data on a multimodal distribution and can be used to inject information as prior knowledge to the generator.
\\
To date, all proposed solutions have been published with the aim of adhering to real data privacy regulations and preventing data leakage during data sharing or the generation of synthetic data for data imputation and augmentation. In contrast, in AQP applications, it is necessary to generate realistic data rather than synthetic data that is as close to real data as possible. The challenges associated with generating tabular data using GANs have been addressed in a few publications since 2017. The purpose of this section is to introduce promising variants of GANs for tabular data generation, followed by a classification of the proposed solutions based on the previously discussed synopsis construction challenges.\\
Choi et al. \cite{choi2017generating} proposed the medical Generative Adversarial Network (medGAN) to generate realistic synthetic patient records based on real data as inputs to protect patient confidentiality to a significant extent. The medGAN generates high-dimensional, multi-label discrete variables by combining an autoen
coder with a feedforward network, batch normalization and shortcut connections. With an autoencoder, flow gradients are able to end-to-end fine-tune the system from discriminator to decoder for discrete patient records. The medGAN architecture uses MSE loss for numerical columns and Cross-Entropy loss for binary columns, and ReLU activation function for both encoder and decoder networks. The medGAN uses the pre-trained autoencoder to generate distributed representations of patient records rather than directly generating patient records. In addition, it provides a simple and efficient method of dealing with mode collapse when generating discrete outputs using minibatch averaging.
\begin{figure}[ht]
    \fbox{\includegraphics[width=\dimexpr\textwidth-2\fboxrule-2\fboxsep]{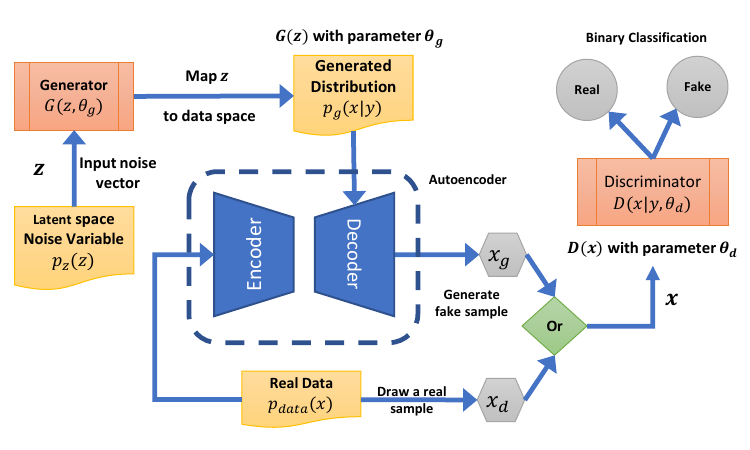}}
    \centering
    \caption{medGAN architecture: Discriminator utilizes autoencoder (which is learned by real data) to receive decoded random noise variable}
    \label{fig:medGan}
\end{figure}
Figure \ref{fig:medGan} shows medGAN architecture and defines the autoencoder role in training process.\\
The generator cannot generate discrete data because it must be differentiable.  To address this issue Mottini et al. \cite{mottini2018airline} proposed a method for generating realistic synthetic Passenger Name Records (PNRs) using Cramer GANs, categorical feature embedding, and a Cross-Net architecture for the handling of this issue (categorical or numerical with null values). As opposed to simply embedding the most probable category, they used the weighted average of the embedded representation of each discrete category. The embedding layer is shared by the generator and discriminator, resulting in a fully differentiable process as a result of this continuous relaxation. For handling null values, they are substituted with a new category in categorical columns. However, continuous columns fill null values with a random value from the same column and then a new binary column is inserted with 1 for filled rows and 0 otherwise. These additional binary columns are encoded like category columns. It should be noted that in this architecture, both the generator and discriminator consist of fully connected layers and cross-layers. Also, except for the last layer (Sigmoid), all layers of the generator use leaky ReLU activations for numerical features and Softmax for categorical features. However, discriminator uses leaky ReLU activations in all but the last layer (linear). Neither batch normalization nor dropout is used in this architecture like Wasserstein and Cramer GANs \cite{bellemare2017cramer}. Data pre-possessing in this algorithm is depicted in Figure \ref{fig:MatrixEmbedding}.
As indicated, discrete values will be embedded using the embedding matrix, followed by the concatenation of them with continuous columns of input data.\\
Table-GAN \cite{park2018data} uses GANs to create fake tables that are statistically similar to the original tables but are resistant to re-identification attacks and can be shared without exposing private information. Table-GAN supports both discrete and continuous columns and is based on Deep Convolutional GANs (DCGANs) \cite{radford2015unsupervised}. Besides the generator and discriminator with multilayer convolutional and deconvolutional layers, the table-GAN architecture also includes a classifier neural network with the same architecture as the discriminator. However, it is trained using ground-truth labels from the original table to increase the semantic integrity of the generated records. Information loss and classification loss are two additional types of loss that are introduced during the backpropagation process. The purpose of these functions is to maintain a balance between privacy and usability while ensuring the semantic integrity of the real and generated data. The information loss compares the mean and standard deviation of real and generated data to measure the discrepancy between them and determine whether they have statistically the same features from the perspective of the discriminator or not, and the classification loss measures the difference between how a record is labeled and how the classifier predicts it should be labeled.
Figure \ref{fig:tableGAN} is a representation of the loss functions in the table-GAN architecture. \\
Xu et al. developed TGAN \cite{xu2018synthesizing}, which is a synthetic tabular data generator for data augmentation that can take into account mixed data types (continuous and categorical). TGAN generates tabular data, column by column, using a Long-Short Term Memory (LSTM) network with attention. The LSTM will generate each continuous column from the input noise in two steps. First, it generates a probability that the column comes from mode $m$, and then normalizes the column value based on this probability. TGAN penalizes GAN's original loss function by adding two KL-divergence terms between generated and real data for continuous and categorical columns separately. Therefore, generator will be optimized as follow:
\begin{equation}\label{eq-TGAN}
\begin{aligned}[b]
\mathcal{L}_G=-\mathbb{E}_{z\sim \mathcal{N}(0,1)}[\log(D(G(z)))]+\sum_{i=1}^{N_c}KL(u^\prime_i,u_i)+\sum_{i=1}^{N_d}KL(d^\prime_i,d_i)
\end{aligned}
\end{equation}
Where $u^\prime_i$ and $u_i$ are probability distribution over continuous column $c_i$ for generated and real data, respectively, $d^\prime_i$ and $d_i$ are probability over categorical column $d_i$ using softmax function for generated and real data respectively, $N_c$ is number continuous columns, and $N_d$ is number of categorical columns. They also also proposed a conditional version of TGAN, named CTGAN \cite{xu2019modeling} , for addressing data imbalance and multimodal distribution problems by designing a conditional generator with training by sampling strategy to validate the generator output by estimating the distance between the conditional distribution over generated and real data.
\\
CTAB-GAN \cite{zhao2021ctab} were introduced with the ability to encode the mixed data type and skewed distribution of input data table, utilizing conditional generator, information and classification loss functions derived from table-GAN, as well as CNNs for both generator and discriminator functions. Since CNNs are effective at capturing the relationship between pixels within an image, therefore they can be employed in enhancing the semantic integrity of created data. However, in order to prepare data tables for feeding CNN, rows are transformed into the nearest square $d \times d$ matrix, where $d=Ceil(\sqrt{N_c+N_d})$, $N_c$ and $N_d$ are number of continuous and categorical columns respectively in a row of the data table, and then, the extra cells values ($d\times d-(N_c+N_d)$) are padded with zeros.\\
It is difficult for GANs to control the generation process of data-driven systems; therefore, integrating prior knowledge about data relationships and constraints can assist the generator in generating synopses that are realistic and meaningful. In order to implement this, DATGAN \cite{lederrey2022datgan} incorporates expert knowledge into GANs generator by matching the generator structure to the underlying data structure using a Directed Acyclic Graph (DAG). Using a DAG, the nodes represent the columns of a data table, while the directed links between them allow the generator to determine the relationship between variables so that one column's generation influences another. It means if two variables have no common ancestors, they will not be correlated in the generated dataset. In relational databases, there is no particular order in which columns appear in data tables. Nevertheless, the DAG enables data tables to have a specific column order based on their semantic relationship.
\subsection{Tabular GAN Evolution}
\begin{figure}[ht]
    \fbox{\includegraphics[width=\dimexpr\textwidth-2\fboxrule-2\fboxsep]{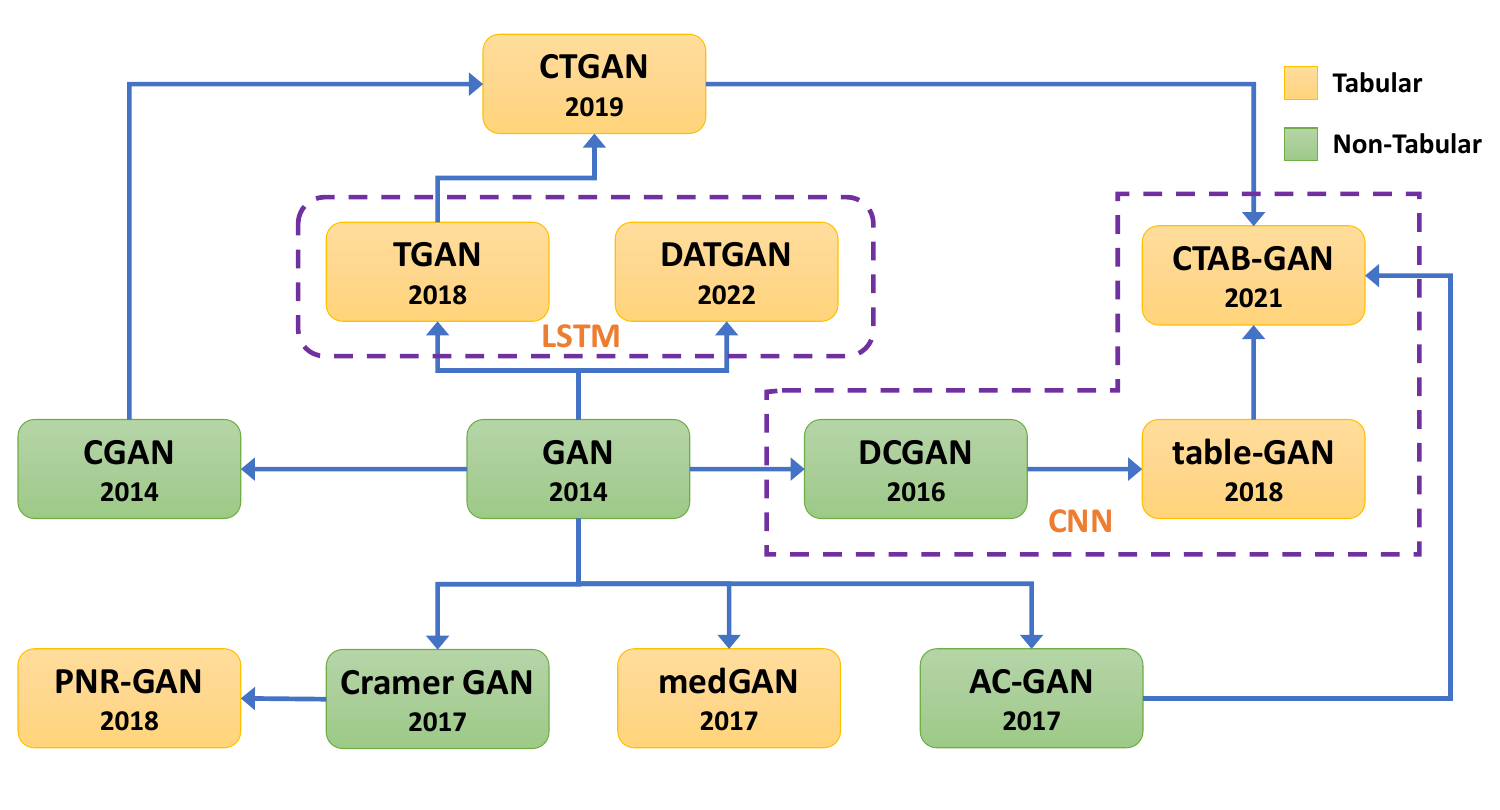}}
    \centering
    \caption{Tabular GAN-based generators evolution based on their relationship. Yellow boxes are tabular generators, and green boxes introduced for non-tabular data.}
    \label{fig:GAN-Taxonomy}
\end{figure}

GAN has made significant progress in recent years, which has led to the development of novel variants that improve on previously introduced versions that had promising results prior to their introduction. Table \ref{tab:GAN-variant} provides a summary of the variants of GAN that have been discussed in this paper. Also, figure \ref{fig:GAN-Taxonomy} shows tabular GAN evolution, along with the year that they were introduced and their ancestors. As shown in this figure, table-GAN and CTAB-GAN utilize convolutional layers as part of their generator, however, CTAB-GAN makes use of a conditional version of generator built on CGAN and AC-GAN. CTGAN also utilizes the conditional version of GAN. With conditioned generators, realistic data can be generated based on the constraints on the data table. On the other hand, TGAN and DATGAN use LSTM for memorizing the data relationships and correlations. Indeed, both conditional generators and LSTMs attempt to generate data based on a prior knowledge about the relationship between columns in a data table.
\begin{table}[hbt!]
\caption{Different tabular GAN architecture and capability.}
\label{tab:GAN-variant}
\renewcommand{\arraystretch}{1.5}
\resizebox{\textwidth}{!}{\begin{tabular}{l|p{3cm}|p{1.7cm}|p{1.8cm}|p{2cm}|l}
\multicolumn{1}{c}{Variant} & \multicolumn{1}{|c}{Capability}& Generator& Discriminator& \multicolumn{1}{|c}{Extra Loss Functions}                     & \multicolumn{1}{|c}{Additional Networks}\\ 
\hline
medGAN& \begin{tabular}[c]{@{}p{3cm}@{}}Generate high-dimensional discrete columns\\ Avoid mode collapse\end{tabular}& FNN& FCN& \begin{tabular}[c]{@{}p{2cm}@{}}MSE\\ Cross-entropy\end{tabular}& Autoencoder\\
\hline
PNR-GAN& \begin{tabular}[c]{@{}p{3cm}@{}}Generate discrete columns\\ Handling null values\end{tabular} & cross-layer FCN & cross-layer FCN & Cramer loss& \\
\hline
table-GAN & Increase semantic Integrity & CNN & CNN & \begin{tabular}[c]{@{}l@{}}Information loss\\ Classification loss\end{tabular} & Classifier (MLP)\\
\hline
TGAN& \begin{tabular}[c]{@{}p{3cm}@{}}Learn multimodal distribution.\\  Generate mixed type variables.\end{tabular}& LSTM& FCN& Cross-entropy& \\
\hline
CTGAN& \begin{tabular}[c]{@{}p{3cm}@{}}Learn non-Gaussian and multimodal distribution.\\  Address imbalance discrete column issue.\end{tabular}& FCN& FCN& Wasserstein loss with gradient penalty& \\
\hline
CTAB-GAN& \begin{tabular}[c]{@{}p{3cm}@{}}Generate discrete and mixed-type column\\ Address imbalance discrete column issue.\\ Learn long-tail distribution\end{tabular} & CNN& CNN & \begin{tabular}[c]{@{}l@{}}Cross-entropy\\ Information loss\\ Classification loss\end{tabular} & Classifier (MLP) \\
\hline
DATGAN & \begin{tabular}[c]{@{}p{3cm}@{}}Increase semantic Integrity\\ Increase representativity of imbalance class\end{tabular} & LSTM & FCN & Wasserstein loss with gradient penalty & DAG \\
\hline
\multicolumn{6}{l}{\begin{tabular}[c]{@{}l@{}}FNN: Feed Forward Neural Network\\ FCN: Fully Connected Neural Network\\ CNN: Convolutional Neural Network\\ MLP: Multi-Layer Perceptron\\ LSTM: Long Short-Term Memory
\end{tabular}}                                                               
\end{tabular}}
\end{table}

%% file: UpdatedSection/3-SynopsisChallenges.tex
Traditional methods in synopsis construction have many challenges relating to data type, structure, distribution, and query aggregation functions. Furthermore, synopses provide the most accurate summary using the entire data stream, and it would be inconvenient to retrieve the entire data set in real-time databases as it changes over time. According to the data structure of relational databases, the challenges associated with generating synopses can be categorized into many significant groups \cite{xu2018synthesizing}.
\subsection{Data type}
It is challenging to construct a data synopsis that is representative of the entire data table due to the difference in data types. For instance, different activation functions on output are required for the generative models since relational database tables include numerical, categorical, ordinal, and mixed data types. As an example of a mixed data type, the financial database contains columns for loan debts, where a loan holder may have no debt or debt with a positive value \cite{zhao2021ctab}. In data analysis, it can be defined as categorical data using a step function, but in reality, it is continuous data. In this regard, a data generator must be able to detect these types of data in order to avoid adverse effects on the interpretation of the data. The several types of data used in creating AQP data synopsis are broken down in Table \ref{tab:data-type}. Mentioning that textual data types are not typically utilized in AQP queries and are therefore ignored here. 
\begin{table}[ht]
\centering
\renewcommand{\arraystretch}{1.5}
\caption{The data types that can be used in AQP queries. Can be aggregated (sum, avg, max, min), can be bounded in where conditions, and can be considered as a group to aggregate other columns.}
\label{tab:data-type}
\begin{tabular}{llp{5cm}p{3cm}}
\hline
\multicolumn{3}{c}{Data types} & Possible role in queries \\ \hline
\multirow{3}{*}{Numerical} & Continuous & Numeric intervals on the real number without finite set of values & aggregation, condition \\
 & Discrete & Finite, countable set of integer number & aggregation, condition, groupby \\ 
 & Mixed & Numeric, but considered as categorical based on the different range & aggregation, condition, groupby \\ \hline
\multirow{3}{*}{Categorical} & Binary & One-hot encoded & condition, groupby \\
 & Textual & One-hot encoding needed & condition, groupby \\
 & Numeric & Treats like textual and numbers are meaningless. & condition, groupby \\ \hline
Ordinal & Numeric & Numeric categories with a clear ordering (like 1-5 rating) & aggregation, condition, groupby\\
\hline
\end{tabular}
\end{table}
\subsection{Bounded Continuous Columns}
Continuous Column $C_i$ is bounded if there are two numbers $a$ and $b$ in which for all $x\in X$ $a\leq x \leq b$. The intricacy in Synopsis construction for these bounded continuous columns arises from the necessity to sample from a probability distribution that not only mirrors the true underlying statistical characteristics of the original data but also adheres strictly to these boundary constraints. For instance, a credit card's expenditure column might be restricted from zero to an upper limit reflecting the credit limit, hence the synthetic data generation process must respect these limits to produce meaningful and applicable synthetic transactions. The challenge intensifies as it requires the synthesis process to be sensitive to the distribution's tails, avoiding the generation of outliers that fall outside the established bounds, which would otherwise lead to unrealistic and operationally irrelevant data points.
\subsection{Non-Gaussian Distribution}
When dealing with the non-Gaussian distributions that are common in real-world datasets, the assumption of normality often fails in the field of synopsis construction. Such distributions may be multimodal, containing several peaks or modes, which reflects the complexity of underlying data-generating processes. For instance, the distribution of incomes in a socio-economic dataset could exhibit multiple modes, corresponding to different socio-economic classes. Traditional synopsis generation techniques may inadequately capture the multi-modes structure of such distributions, leading to the missing of entire modes. This results in a generated synopsis that fails to represent segments of the population within the original dataset \cite{xu2018synthesizing}.\\
Moreover, the presence of long-tailed distributions poses additional challenges \cite{zhao2021ctab}. These distributions are characterized by a proliferation of infrequent events, such as a customer purchase history where a vast majority of customers make infrequent purchases, while a minor fraction exhibits high purchase frequencies. Synthesizing data from such a distribution requires not only capturing the frequent low-occurrence events but also accurately representing the rare high-occurrence instances. The conventional methods may struggle with this, often either over-representing the tail and creating too many rare events or under-representing it, thus failing to capture the true nature of the underlying data. This misrepresentation can skew the synopsis, rendering it less effective for use in decision-making processes where an understanding of rare events is critical.
\subsection{Imbalance Categorical Column}
In the construction of data synopses, the handling of imbalanced categorical columns presents a significant challenge \cite{xu2018synthesizing}. Categorical variables in real-world datasets frequently show a skewed distribution in terms of the frequency of occurrence across categories. The presence of such a disparity indicates that minority categories make only a small contribution to the overall distribution of data, which may result in their under-representation in the generated synopsis. The process of creating synopses is influenced by a lack of representation of certain classes, resulting in a bias towards the majority class due to its higher statistical likelihood. For instance, consider a customer gender column in a retail database with a pronounced imbalance, where 'male' customers vastly outnumber 'female' customers. A synopsis generated from this distribution might reflect this skew, resulting in a synthetic dataset dominated by 'male' entries. However, this skew inaccurately portrays the significance of the 'female' category, which, despite its smaller size, may carry substantial weight in consumer behavior analysis.
\begin{figure}[ht]
    \fbox{\includegraphics[width=\dimexpr\textwidth-2\fboxrule-2\fboxsep]{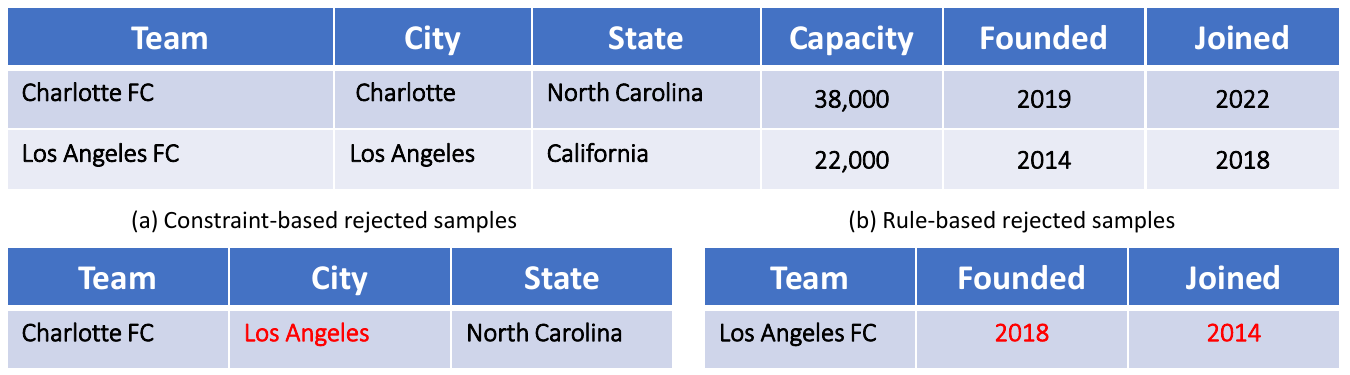}}
    \centering
    \caption{Each soccer team in the table corresponds to a particular location and has a specific capacity, foundation year, and year of entry into MLS. (a) and (b) shows two examples of constraint-based and rule-based sample rejection during the data synopsis generation process.}
    \label{fig:semantic-example}
\end{figure}
\subsection{Semantic relationship and Constraint}
Tabular data often includes complex semantic relationships that are not easily understood through standard statistical analysis \cite{khurana2020semantic}. These relationships can exist between categorical and numerical columns alike and are crucial for maintaining the integrity and usefulness of the generated synopsis. Identifying and encoding such relationships is a challenge due to the heterogeneity of domain-specific constraints and the complex nature of the inter-column dependencies which may not be amenable to simple rule-based generalizations. For instance, semantic relationships may determine that certain numerical values possess validity solely when paired with specific categorical entries, imposing a constraint-based association. Alternatively, a rule-based linkage could suggest a probabilistic co-occurrence pattern between different fields in the data. Hence, it is imperative for a comprehensive process of generating synopses to include mechanisms that can deduce these complex relationships, which can be multi-faceted and deeply embedded within the structure of the data. A failure to do so not only compromises the authenticity of the synthesized data but also limits the operational relevance of the synopsis, as it could lead to the generation of implausible or inconsistent records that do not adhere to the real-world rules and constraints governing the dataset. Figure \ref{fig:semantic-example} represents two examples of generated samples from a table that the model should reject semantically. To generate a representative data synopsis in AQP, the city must be properly associated with the state, and the joined column cannot precede the founded column.

%% file: UpdatedSection/4-Solution.tex
It is possible to categorize synopsis construction solutions into three different categories: Data Transformation, which addresses data type issues; Distribution Matching, which addresses ranges and distributions of data; and Conditional and Informed Generator, which addresses imbalance classes, semantic relationships, and table constraints.
\subsection{Data Transformation}
\begin{figure}[ht]
    \fbox{\includegraphics[width=\dimexpr\textwidth-2\fboxrule-2\fboxsep]{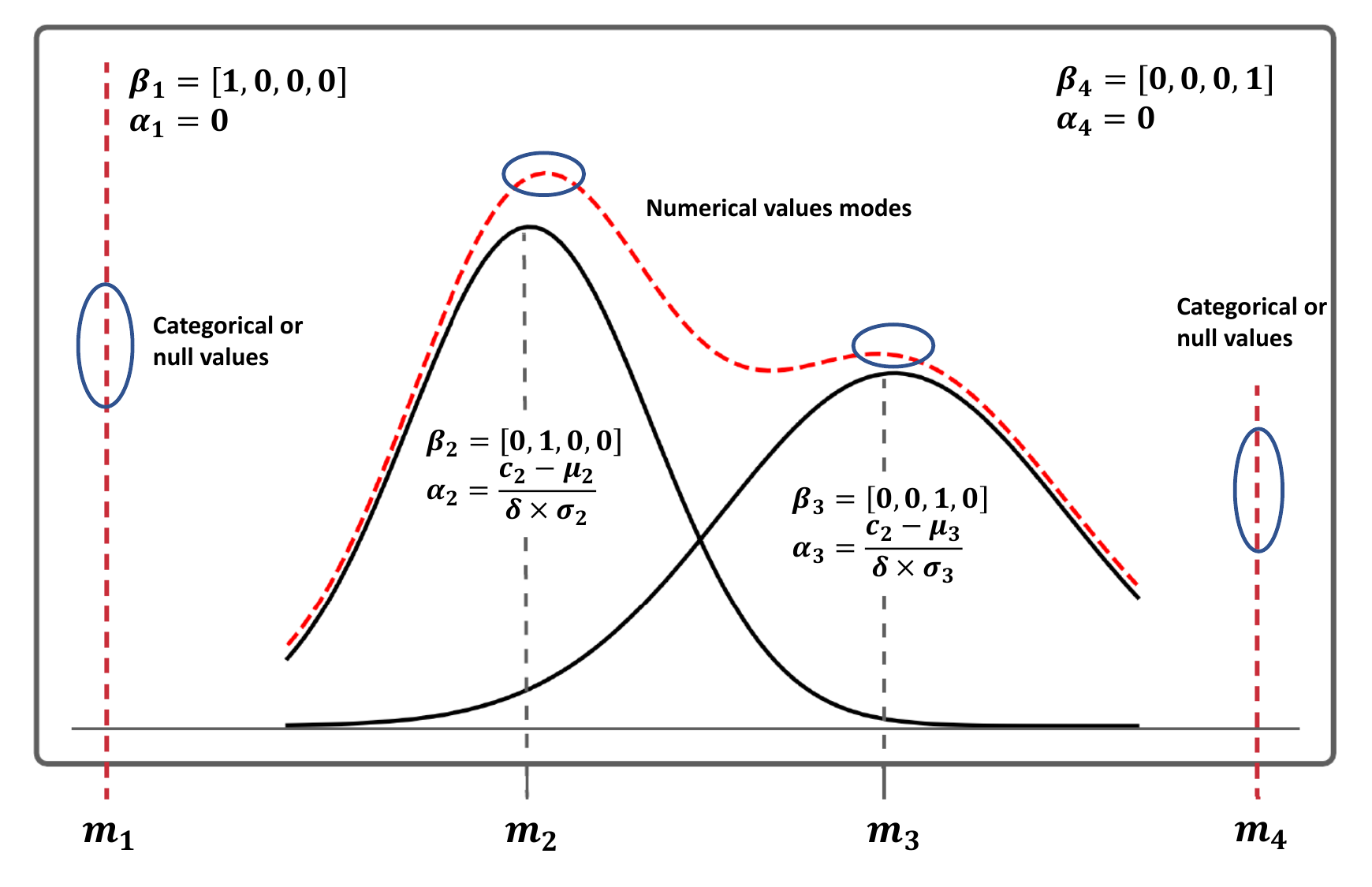}}
    \centering
    \caption{Distribution over a mixed-type column. $m_1$ and $m_4$ represent the categorical part or null values of this column, whereas $m_2$ and $m_3$ represent modes for numeric parts. The numeric parts are defined by Variational Gaussian Mixture (VGM) model.
 \cite{zhao2021ctab}}
    \label{fig:mixed-mode}
\end{figure}

 Mode normalization is capable of detecting modes of data by assigning samples to different modes and then normalizing each sample based on the corresponding mode estimator \cite{deecke2018mode}. To deal with multimodal distribution for continuous columns, mode-specific normalization is introduced in TGAN \cite{xu2018synthesizing}. Using this algorithm, first, the number of continuous columns' modes is calculated using Gaussian kernel density estimation. Then, Gaussian Mixture Model (GMM) can be employed to efficiently sample values from a distribution with multiple modes by clustering the values of continuous columns ($C_i$). In other words, the weighted sum of the Gaussian distributions over $C_i$ can represent the multimodal distribution over it. A normalized probability distribution over $m$ Gaussian distributions can then be used to represent each continuous column so that each column can be clustered into $m$ fixed Gaussian distributions. As a result, if there are less than m modes in one column, then the probability of that mode is high, and for the rest, it is close to zero. However, in CTGAN \cite{xu2019modeling}, first, a Variational Gaussian Mixture model (VGM) should be applied to each continuous column ($C_i$) in order to fit a Gaussian mixture and find the number of modes ($m$). Then, a one-hot vector ($\beta_{i,j}$) indicates to which mode a given value belongs, and a scalar ($\alpha_{i,j}$) serves as the value itself within that mode. For the learned Gaussian mixture for column $C_i$ with $m$ modes, the following equation is given:
\begin{equation}\label{eq2}
\begin{aligned}[b]
\mathbb{P}_{C_i}(c_{ij})=\sum_{k=1}^{m}w_k\mathcal{N}(c_{ij};\mu_k,\sigma_k)
\end{aligned}
\end{equation}
where $c_{i,j}:$ value of $j^{th}$ row from $i^{th}$ column, $\mu_k$ and $\sigma_k:$ the mean and standard deviation of Gaussian distribution for $k^{th}$ mode, and $w_k$ is the weight of $k^{th}$ mode. For each value, the probability density $\rho$ of  $k^{th}$ mode is:
\begin{equation}\label{eq:prob-density}
\begin{aligned}[b]
\rho_k=w_k\mathcal{N}(c_{ij};\mu_k,\sigma_k)
\end{aligned}
\end{equation}
Therefore, each value can be normalized according to the mode with highest probability. As an example, the values of $\alpha$ and $\beta$ related to column $c_{i,j}$ in $k^{th}$ mode will be:
\begin{equation}\label{eq:alpha-beta}
\begin{aligned}[b]
\alpha_{i,j}=\frac{c_{i,j}-\mu_k}{\delta\sigma_k}\;,\;\;\; \beta=[0,0,1,...,k]
\end{aligned}
\end{equation}
where $\delta$ is a parameter specified by the modeller.\\
For categorical columns $D$, the situation is different; TGAN \cite{xu2018synthesizing} stated to convert these columns ($d_{ij}$) to a representation using one-hot encoding with added noise ($Uniform(0,\gamma)$, $\gamma$ is an arbitrary number). To achieve this, after creating the one-hot vector, noise will be added to each element, and the resulting representation will be renormalized. Therefore, each data row can be represented by a concatenation of continuous and categorical columns as follows:
\begin{equation}\label{eq16}
\begin{aligned}[b]
row_j=\alpha_{1,j}\oplus\beta_{1,j}\oplus ... \oplus\alpha_{N_c,j}\oplus\beta_{N_c,j}\oplus d_{1,j}\oplus ... \oplus d_{N_d,j}
\end{aligned}
\end{equation}
where $d_{i,j}$ is one-hot representation of a categorical column, $N_c$ is number of continuous columns and $N_d$ is number of categorical columns $D_i$.\\
As previously discussed, columns can be considered mixed if they contain both categorical and continuous values or continuous values with null values. The encoding process for continuous and categorical columns in CTAB-GAN \cite{zhao2021ctab} is exactly the same as CTGAN \cite{xu2019modeling} by defining $\alpha$ and $\beta$. However, in mixed-type columns, the encoder is defined so that each column is considered a concatenation of value-mode pairs, where the categorical part of values takes zero for $\alpha$ and is treated as continuous. Figure \ref{fig:mixed-mode} shows the distribution over an arbitrary mixed-type column, with two modes for continuous ($m_2,m_3$) and two categorical parts ($m_1,m_4$) and illustrates how this algorithm transforms one row of mixed-mode data.
\subsection{Distribution Matching}
In order to generate synopses with the same distribution as the underlying distribution, the training algorithm should penalize the generator. Information loss \cite{park2018data} helps generator to generate synopses statistically closer to the real one. It utilizes the statistical characteristics $\mathcal{L}_{mean}$ (first-order statistics, Eq. \ref{eq11}) and $\mathcal{L}_{sd}$ (second-order statistics Eq. \ref{eq12}) of the extracted features prior to the classifier in the discriminator to penalize the generator for the discrepancy between real and generated data. This makes sense because the extracted features are used to determine the binary decision of the discriminator.
\begin{equation}\label{eq11}
\begin{aligned}[b]
\mathcal{L}_{mean} = \parallel \mathbb{E}[\boldsymbol{f}_x]_{x\sim p_{data}(x)}-\mathbb{E}[\boldsymbol{f}_{G(z)}]_{z\sim p_z(z)}\parallel_{2}
\end{aligned}
\end{equation}
\begin{equation}\label{eq12}
\begin{aligned}[b]
\mathcal{L}_{sd} = \parallel \mathbb{SD}[\boldsymbol{f}_x]_{x\sim p_{data}(x)}-\mathbb{SD}[\boldsymbol{f}_{G(z)}]_{z\sim p_z(z)}\parallel_{2}
\end{aligned}
\end{equation}
Where $\boldsymbol{f}$ represents features, $\mathbb{E}[\boldsymbol{f}]$ is the average and $\mathbb{SD}[\boldsymbol{f}]$ is the standard deviation of features over all rows in the data table. The Euclidean norm is used to measure the discrepancy between two terms. As we discussed before, table-GAN \cite{park2018data} was developed to protect confidential data privacy when it is shared with the public. As a result, it should be possible to control the similarity of generated data with real data during the generating process. To this end, information loss for the generator is demonstrated as follows:
\begin{equation}\label{eq13}
\begin{aligned}[b]
\mathcal{L}_{info}^G = \max(0,\mathcal{L}_{mean}-\delta_{mean})+\max(0,\mathcal{L}_{sd}-\delta_{sd})
\end{aligned}
\end{equation}
Where $\delta$ is a threshold indicating a quality degradation of generated data and $\max(.)$ represents the hinge-loss that is zero until $\delta$ is reached. However, in AQP, it is not necessary to meet this threshold in order to generate realistic data synopses.\\
DATGAN \cite{lederrey2022datgan} uses the improved version of the Wasserstein loss function in WGAN \cite{arjovsky2017wasserstein} in addition to the Vanilla GAN loss function with gradient penalty \cite{gulrajani2017improved} and also add the KL-divergence as an extra term to the original loss function. Both of these terms aim to minimize the difference between the probability distributions of real and generated data. WGAN employs an alternative method of training the generator to better approximate real data distribution. This approach replaces the discriminator model with a critic that scores the degree to which a data sample is real or fake rather than using the discriminator as a classifier. Therefore, WGAN considers discriminator output as a scalar score instead of a probability, and Wasserstein loss ensures a greater difference between the scores for real and generated data. As a result, it can prevent vanishing gradients in the generator models. However, the WGAN's primary problem is that it must clip the weights of the critic in order to enforce the Lipschitz constraint. This issue can be addressed by adding a gradient penalty to the critic. Eq. \ref{eq6} shows the Wasserstein objective function, and Eq. \ref{eq7} shows the same with a penalty on the gradient norm for random samples $\hat{x}\sim p_{\hat{x}}$.
\begin{equation}\label{eq6}
\begin{aligned}[b]
\underset{G}{\mathrm{min}}\ \underset{D}{\mathrm{max}}V(D,G)=\mathbb{E}_{x\sim p_{data}(x)}[D(x)]-\mathbb{E}_{z\sim p_z(z)}[D(G(z))]
\end{aligned}
\end{equation}
\begin{equation}\label{eq7}
\begin{aligned}[b]
L_{W}=\mathbb{E}_{z\sim p_z(z)}[D(G(z))]-\mathbb{E}_{x\sim p_{data}(x)}[D(x)]+\lambda\:\mathbb{E}_{\hat{x}\sim p_{\hat{x}}}[||\nabla_{\hat{x}}D(\hat{x})||_2-1)^2]
\end{aligned}
\end{equation}
where $\lambda$ is a parameter defined by the modeler and $\hat{x}$ sampled from $G(z)$ and $x$.
\subsection{Conditional and Informed Generator}
\begin{figure}[ht]
    \fbox{\includegraphics[width=\dimexpr\textwidth-2\fboxrule-2\fboxsep]{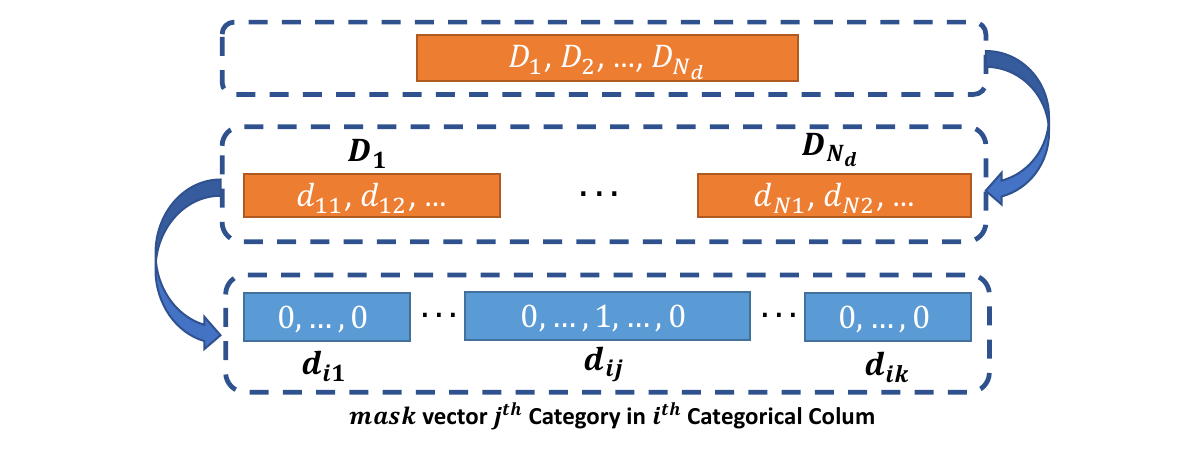}}
    \centering
    \caption{Following vectorization of categorical columns, all vectors will be initiated by 0, then $j^{th}$ category from $i^{th}$ column will be selected, and the value of the corresponding element will be changed to 1.}
    \label{fig:mask-vector}
\end{figure}

\begin{figure}[ht]
    \fbox{\includegraphics[width=\dimexpr\textwidth-2\fboxrule-2\fboxsep]{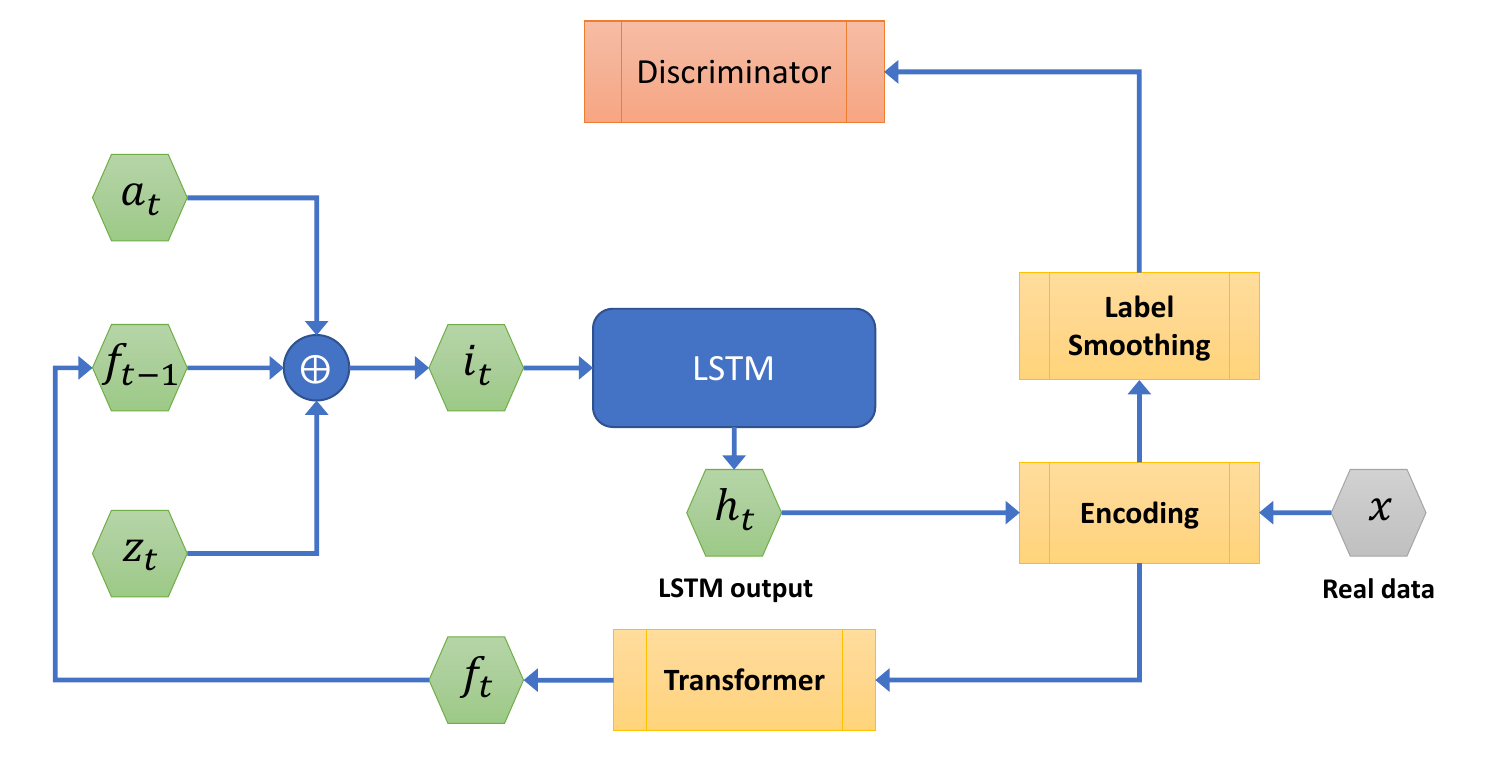}}
    \centering
    \caption{DATGAN process flow digram.}
    \label{fig:lstm}
\end{figure}

Imbalances in categorical columns can cause inaccuracies when generating synopses and may result in the generator not being trained to match the distribution of the real data. In CTGAN \cite{xu2019modeling}, the conditional generator is introduced (using training-by-sampling) as a solution to this problem. To this aim, the generated value can be interpreted as a conditional distribution of rows given the value of an imbalanced categorical column. Therefore, the original distribution can be reconstructed as follows:
\begin{equation}\label{eq3}
\begin{aligned}[b]
P_g(row|D_i=k)=P(row|D_i=k)\; => \; P(row)=\sum_{k\in{D_i}}P_g(row|D_i=k)P(D_i=k)
\end{aligned}
\end{equation}
where $k$ is a value in $i^{th}$ categorical cloumns $D_i$. For the implementation of this solution, a conditional vector consisting of a mask vector that represents the address of the table value (column and corresponding row value) is required. This conditional vector does not guarantee the feed-forward pass obtains the correct value based on the mask vector $M$; instead, the suggested approach penalizes the conditional generator's loss by averaging the cross-entropy between the generated $\hat{M}_i$ and the expected conditional vector $M_i$ over all instances of the batch. The generator loss can be expressed as follows:
\begin{equation}\label{eq4}
\begin{aligned}[b]
\mathcal{L}_G=\mathbb{E}[H(M_i,\hat{M}_i)]
\end{aligned}
\end{equation}
Where H(.) is cross-entropy between two values. As a result, the generator learns to replicate the masked value in the generated row during training. The conditional vector for a data table with $N$ categorical columns is the direct sum of all mask vectors ($M$) across each column $D_i$, where for each value $c_{i,j}:$
\begin{equation}\label{eq41}
\begin{aligned}[b]
M_i = \left\{
    \begin{array}{lr}
        1 & \text{if } j^{th} \text{value}\\
        0 & \text{the rest}
    \end{array}
\right\}\text{, } cond=M_1\oplus\ldots\oplus M_N
\end{aligned}
\end{equation}
In fact, generator loss allows the generator to learn to produce the same classes as the given conditions. Mask vectors ($M_i$) are initialized with 0 for each categorical column ($D_i$) during the conditional generator procedure. Then, a column is chosen at random, and the Probability Mass Function (PMF) is applied to the column's range of categories. According to PMF, one category is then picked, and its value in the corresponding mask vector is changed to 1. Finally, the conditional vector is formed, and the generator is able to generate a synthetic row for the given categorical column. Figure \ref{fig:mask-vector} represents a mask vector generation process for a data table with $N_d$ categorical columns when generator is conditioned for $j^{th}$ category of $i^{th}$ categorical column.\\
It has been discussed previously that columns in a table may have a meaningful relationship with one another. CTAB-GAN \cite{zhao2021ctab} utilizes a classifier neural network using auxiliary classifier GAN (AC-GAN) \cite{odena2017conditional} that is a conditional GAN type that requires the discriminator to predict the class label $c\sim p_c$ of generated data as well as the realness classifier. In AC-GAN, the generator generates a new sample using noise $z$ and a class label $c$, while the discriminator provides both a probability distribution over sources $P(S|X)$ and a probability distribution over class labels $P(C|X)$. The objective function contains the following terms:
\begin{equation}\label{eq14}
\begin{aligned}[b]
\mathcal{L}_S=\mathbb{E}_{x\sim p_{data}(x)}[\log P(S=real)]+\mathbb{E}_{x\sim p_z(z)}[\log P(S=fake)]
\end{aligned}
\end{equation}
\begin{equation}\label{eq15}
\begin{aligned}[b]
\mathcal{L}_C=\mathbb{E}_{x\sim p_{data}(x)}[\log P(C=c_{real})]+\mathbb{E}_{x\sim p_z(z)}[\log P(C=c_{fake})]
\end{aligned}
\end{equation}
Where $\mathcal{L}_S$ is likelihood of predicting the correct source, $\mathcal{L}_C$ is likelihood of predicting the correct class, and c is a class label. Discriminator is trained to maximize $\mathcal{L}_C + \mathcal{L}_S$ and generator is trained to maximize $\mathcal{L}_C - \mathcal{L}_S$. These objective functions allow the training procedure to generate data according to a specific type of data, while the discriminator must predict the class label of the generated data and determine whether or not it is real. As a result of this, the classifier loss (Eq. \ref{eq:classifier-loss}) will be added to the generator in CTAB-GAN to increase the semantic integrity of generated records and penalizes generator where the combination of columns in a data row is semantically incorrect.
\begin{equation}\label{eq:classifier-loss}
\begin{aligned}[b]
\mathcal{L}_{class}^G=\mathbb{E}_{z\sim p_z(z)}[|l(G(z))-C(fe(G(z)))|]
\end{aligned}
\end{equation}
where $l(.)$ returns the target label and $fe(.)$ returns the input features of a given row.\\
As mentioned before, DATGAN \cite{lederrey2022datgan} uses DAG to control the generation process based on semantic relationships and correlations between columns. According to the constructed DAG, each column and its sequence are represented by Long Short Term Memory (LSTM) cells. Therefore, by providing the generator with prior knowledge, DAG decreases the GAN's capacity to overfit noise in the training process and enables the GAN to produce more accurate data by using these noises more efficiently. Inputs and outputs of LSTM cells should be modified in accordance with the GAN architecture. Inputs can be expressed as follows:
\begin{equation}\label{eq5}
\begin{aligned}[b]
i_t=a_t\oplus f_{t-1} \oplus z_t
\end{aligned}
\end{equation}
where $z_t$ is a tensor of Gaussian noise, which is the concatenation of the noise from the source nodes at each node of the DAG. $f_{t-1}$ is the transformed output of previous tensor ($h_{t-1}$). For the purposes of determining which previous cell outputs are relevant to a node input, $a_t$ represents a weighted average of all ancestor LSTM outputs. Therefore, $a_t$ and the $z_t$ are defined based on all ancestors of the current node. Data input into DATGAN architecture (generated and real data) should be encoded into [-1,1] or [0,1] using techniques described in the "Data Transformation" section. Additionally, for categorical columns, generators produce probability over each class, making it easy for a discriminator to differentiate between real and created values. Therefore, DATGAN recommends using one-sided label smoothing for the default loss. It means the categorical {0,1} vectors are introduced with additive uniform noise and then rescaled to [0,1] bound vectors. Figure \ref{fig:lstm} illustrates the DATGAN process flow diagram, including the data transformer and label smoothing.
In this algorithm, DAG is generated manually; therefore, semantic relationships between variables should be injected as expert knowledge and cannot be detected by the model. However, tabular data cannot be considered sequential since the order of columns in a data table is generally random. Therefore, a DAG is used to create a specific sequence of columns.

%% file: sections/9-Evaluation.tex
To avoid performing expensive computations and to take a trial-and-error approach, AQP requires an estimation of errors before running the query. As a result, the AQP system is able to select the optimum synopsis type and resolution based on the user's latency or accuracy requirements. Error Quantification Modules (EQMs) in AQP systems perform this process by measuring the quality of responses either by predicting or by running queries on synopses \cite{mozafari2015handbook}. A broad classification of the criteria for evaluating approximate query processing systems can be made as follows \cite{dell2010accuracy}:
\begin{itemize}
  \item \textbf{Query type:} In terms of aggregation functions and conditions, what types of queries are covered by methods?
  \item \textbf{Time Complexity:} How long does it take to produce the synopses and return an approximate result?
  \item \textbf{Space Complexity:} What is the required storage space for data synopses?
  \item \textbf{Accuracy:} Does the approximate answer meet the error confidence interval?
\end{itemize}
In this study, however, the focus is on the creation and quality of synopses, therefore, this section discusses the synopses evaluation technique rather than the EQM process. Different aggregation functions necessitate unique considerations when creating a synopsis. For example, Min() and Max() require that the synopsis include the critical outliers which contain critical information. For instance, as depicted in Figure 4, a single transaction in a real-world database may contain vital information and influence decisions based on business problems. As can be seen in this figure, a food supplier sells its product to both retailers and wholesalers. Although retailers make the majority of sale transactions, wholesalers can place orders that are more significant in terms of quantity and cost.
\begin{figure}[ht]
    \includegraphics[width=\dimexpr\textwidth-2\fboxrule-2\fboxsep]{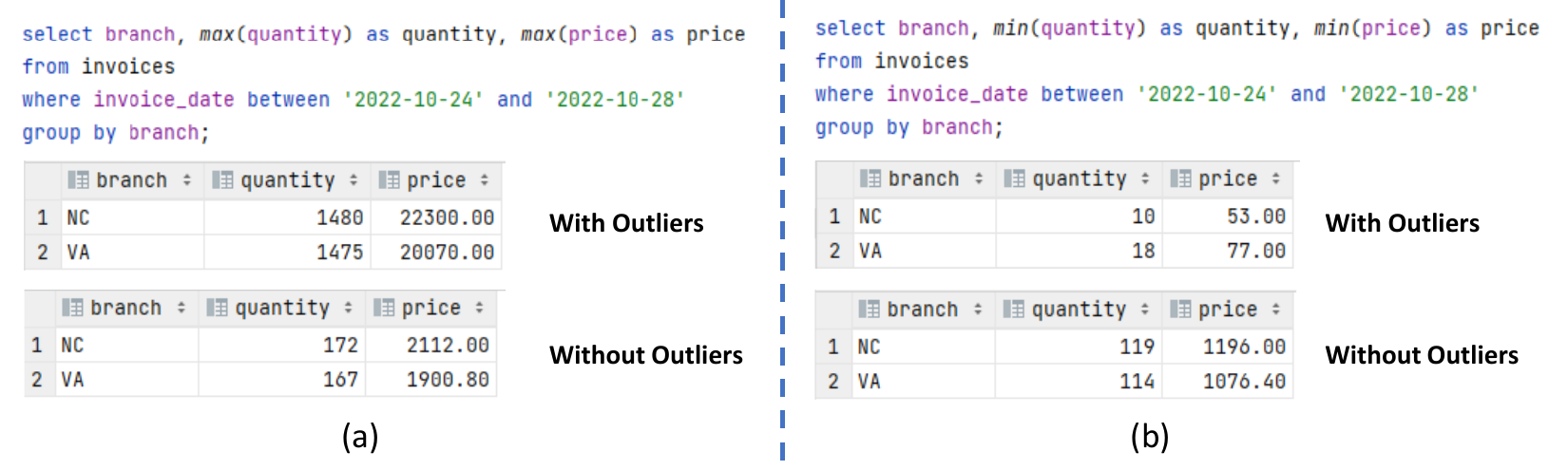}
    \centering
    \caption{In the "invoices" table, there are only two records with outliers for each branch. (a) The aggregated query group by the branch is shown to return the maximum quantity and maximum price sold in a week, and (b) shows a similar query for minimum values. Obviously, approximate results for min() and max() are unreliable when dealing with outliers.}
    \label{fig:outliers}
\end{figure} \\
Likewise, Count() and Sum() may return unacceptable results for minority groups if the query aggregates groups by those columns. Avg() can also be sensitive in situations where a query filters data based on specific conditions or criteria.\\
In generative models, evaluation methods cannot be generalized to other contexts; instead, they must be evaluated explicitly based on their application. In GMs optimization, Gaussian distributions are fitted to a mixture of Gaussian distributions by minimizing distance measures such as Maximum mean discrepancy (MMD) \cite{gretton2006kernel} and Jensen-Shannon divergence (JSD). Minimizing MMD or JSD results in the omission of some modes in a multimodal distribution. In addition, maximizing average log-likelihood or minimizing KL-divergence can assign large probabilities to non-data regions. In image synthesizing applications, three common criteria are used to evaluate generative models: log-likelihood, Parzen window estimates, and visual fidelity of samples \cite{theis2015note}. However, the evaluation of results for tabular data with complex data types and distribution would be quite different.\\
In order to measure accuracy, a generated synopsis should first demonstrate that it is a good representation of real data. The SDMetrics Python library \cite{sdmetrics} introduces a set of metrics to measure the quality and privacy of synthetic data. However, considering data privacy is not the goal of this study and may reduce the quality of the data. These metrics are summarized and reformed to make them suitable for evaluating the generated synopses in AQP. In order to achieve this objective, the comparison of two real and generated datasets can be divided into the following categories: Data Coverage; Data Constraint; Data Similarity, and Data Relationship.\\
\subsection{Data Coverage}
For discrete columns $D_i$, we must determine whether all categories in the real data are represented in the synopsis. To accomplish this goal, A score is calculated by dividing the number of unique categories in the synopsis by the number of unique categories in the corresponding column of the actual data as follow:
\begin{equation}\label{eq-catcoverage}
\begin{aligned}[b]
coverage_{Di}=\left(\frac{N_{D_{g}}}{N_{D_{data}}}\right)_i
\end{aligned}
\end{equation}
Where $i$ is the column index, $N_{D_g}$ is number of unique categories in the generated synopsis and $N_{D_{data}}$ is the number of unique categories in the real data. When a column is scored 1, all of the unique categories in the actual data are present in the generated synopsis, while a score of 0 indicates that no unique categories are present in the generated synopsis. In the case of continuous columns, the coverage metric is used to measure whether a generated column in the synopsis covers the whole range of values that can be found in the real column. The coverage score for continuous columns is calculated as follows:
\begin{equation}\label{eq-concoverage}
\begin{aligned}[b]
coverage_{Ci}=1-\left[max\left(\frac{min(C_{g})-min(C_{data})}{max(C_{data})-min(C_{data})},0\right)+max\left(\frac{max(C_{data})-max(C_{g})}{max(C_{data})-min(C_{data})},0\right)\right]
\end{aligned}
\end{equation}
Where $C_g$ is the generated value and $C_{data}$ is the real value of column $C_i$. The goal of this metric is to determine how closely the min and max of the generated values match the actual min and max values. It is possible for the equation \ref{eq-concoverage} to become negative if the range covered by the generated synopsis is inadequate, and in such a situation, it returns a score of 0 since it is the lowest possible result.
\subsection{Data Constraint}
In order to measure how a continuous column adheres to a boundary of real data, Boundary Adherence is introduced. The frequency of generated values within the minimum and maximum ranges of the real column values is calculated using this metric.  
\begin{equation}\label{eq-BoundaryAdherence}
\begin{aligned}[b]
adherence_{Ci}=\frac{N_{(min<x_i<max)}}{N_i}
\end{aligned}
\end{equation}
Where $N_i$ is the number of records in the column $C_i$. A column with a score of 1 indicates all values adhere to the boundaries of real data, while a column with a score of 0 indicates that no values fall between the minimum and maximum of the real data.
\subsection{Data Similarity}
\begin{figure}[ht]
    \includegraphics[width=\dimexpr\textwidth-2\fboxrule-2\fboxsep]{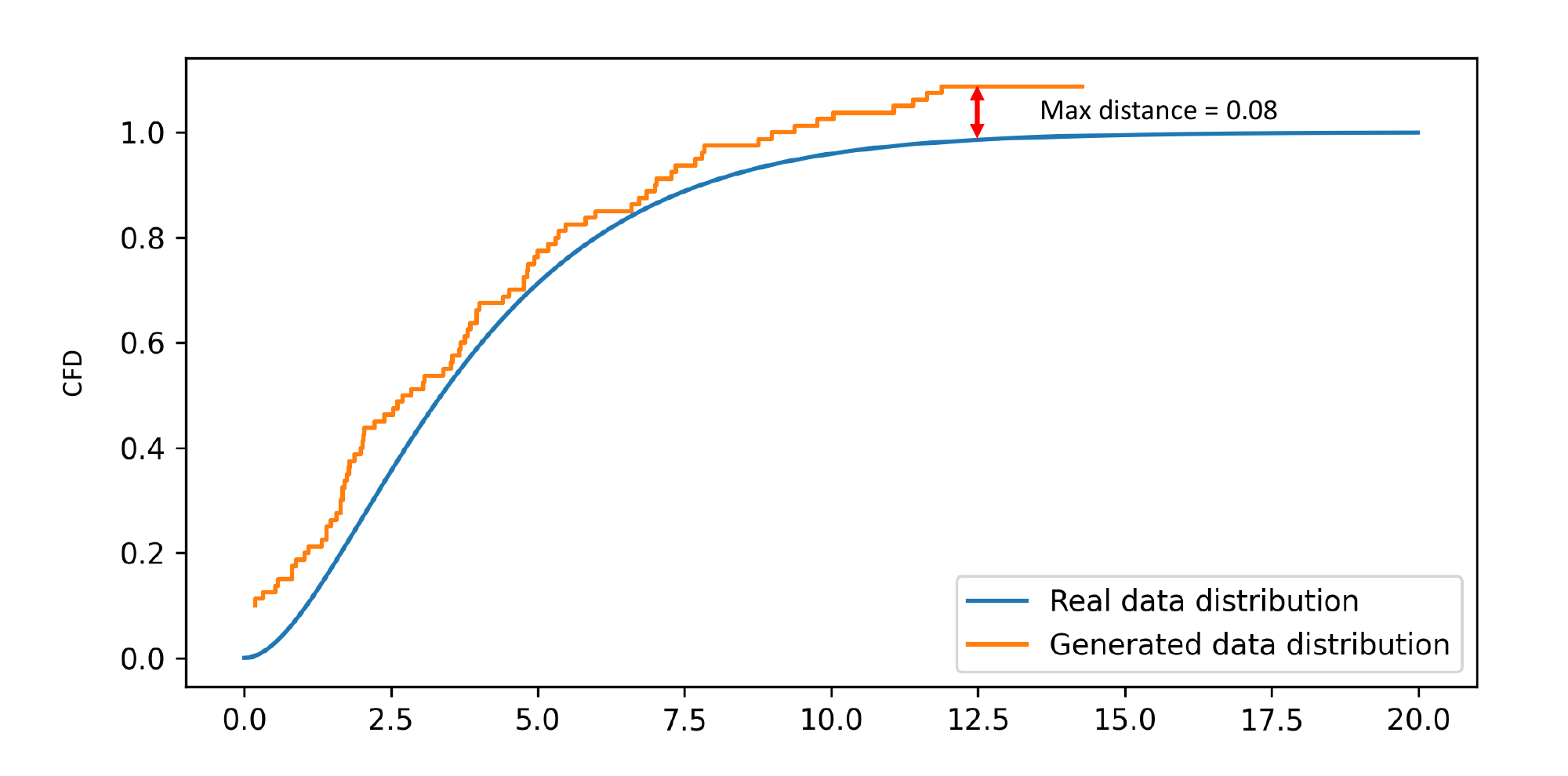}
    \centering
    \caption{Distances are measured between 0 and 1, but the complement of this metric can also be considered. Therefore, a higher score indicates a higher quality according to 1-(KS statistic distance) \cite{sdmetrics}.}
    \label{fig:KSComplement}
\end{figure}

The Synthetic Data Metrics (SDMetrics) library \cite{sdmetrics} introduced several metrics for measuring data similarity. In order to calculate the similarity between real and generated marginal distribution, two types of metrics are available: the Kolmogorov-Smirnov (KS) statistic for continuous columns and the Total Variation Distance (TVD) for discrete columns. Based on the KS statistic, we can determine how much the empirical distribution function of the generated data differs from the cumulative distribution function (CDF) of the real data. This means that, in this case, the KS statistic represents the maximum difference between the two generated and real CDFs, as illustrated in figure \ref{fig:KSComplement}.
 \\
It can be calculated using the following expression:\\
\begin{equation}\label{eq-KS}
\begin{aligned}[b]
KS_{data, g}=\underset{x}{sup}\left|F_{1,data}(x)-F_{2,g}(x)\right|
\end{aligned}
\end{equation}
Where $F_{1,data}$ and $F_{2,g}$ are the empirical distribution functions of the real and generated data, respectively, and $sup$ is the supremum function.\\
In order to calculate the TVD statistic, it first computes the frequency of each category value in the real and generated columns and expresses it as a probability. Once the frequency is calculated, the TVD statistic compares the difference in probabilities using the following formula:
\begin{equation}\label{eq-TV}
\begin{aligned}[b]
TVD_{data,g}=1-\delta(X,G)=1-\frac{1}{2}\sum_{x\in{D_{data}}}|X_{x,g}-G_{x,g}|
\end{aligned}
\end{equation}
Where x and g refer to all possible categories in discrete column D, and X and G represent the frequencies for those categories for real and generated data, respectively. The similarity score is considered the complement of a TVD, so a higher score indicates a higher level of quality.\\
A missing value (null value) can occur in many different situations in a real-world database \cite{biskup1981formal}. This is why the missing value similarity score has been introduced for the purpose of detecting these missing values in generating a synopsis. In this metric, the number of missing values in the generated data is compared to the number of missing values in the real data for each column. The metric can be applied to any column with any type of data. As part of this process, the proportion of missing values in both the real and synthetic data is calculated, and the normalized version of this (equation \ref{eq-MissingValueSimilarity}) results in a similarity score between 0 and 1, with 1 representing the highest level of similarity.
\begin{equation}\label{eq-MissingValueSimilarity}
\begin{aligned}[b]
missing=1-|X_{p}-G_{p}|
\end{aligned}
\end{equation}
Where p is the proportion of missing values and X and G represent the distribution for real and generated data, respectively. In addition, it is possible to measure the statistical similarity between a column of real data and a column of generated data using mean, median, and standard deviation using the following formula:
\begin{equation}\label{eq-StatisticalSimilarity}
\begin{aligned}[b]
similarity=\max\left(1-\frac{|f(x)-f(g)|}{|\max(x)-min(x)|},0\right)
\end{aligned}
\end{equation}
Where an arithmetic mean, median, or standard deviation is defined as f and it returns a score between 0 and 1, where a high value represents a high degree of similarity.\\
\subsection{Data Relationship}
For measuring semantic relationship and correlation between columns within a single data table, Contingency can be applied to discrete columns using crosstabulation (a.k.a contingency table). This score is a matrix representation of the multivariate frequency distribution of variables. First, two contingency tables should be created over the present categories in each column in order to compare a discrete column in the real data with the corresponding column in the generated data. Indeed, the created tables summarize the proportion of rows in real and generated data that have each combination of categories. After that, the total variation distance is used to calculate the difference between the contingency tables. In this case, the distance would be between 0 and 1, so subtracting 1 from the score would indicate a high degree of similarity. Below is a formula that summarizes the process.\\
\begin{equation}\label{eq-contingency}
\begin{aligned}[b]
contingency_{x, g}=1-\frac{1}{2}\sum_{x\in{D_{data}}}\sum_{g\in{D_{g}}}|X_{x,g}-G_{x,g}|
\end{aligned}
\end{equation}
Where x and g refer to all possible categories in discrete column D, and X and G represent the frequencies for those categories for real and generated data, respectively. A score of 1 indicates the best contingency between real and generated data, and a score of 0 indicates the worst contingency. Also, a correlation similarity test can be applied to continuous columns by measuring the correlation between two numerical columns and computing the similarity between the real and generated data using Pearson's and Spearman's rank coefficients. Initially, a correlation coefficient should be calculated between two continuous columns in the real data and their corresponding columns in the generated data. Then, after normalizing two correlation values, the following equation returns a similarity score.
\begin{equation}\label{eq-CorrelationSimilarity}
\begin{aligned}[b]
correlation_{x, g}=1-\frac{|X_{x,g}-G_{x,g}|}{2}
\end{aligned}
\end{equation}
Where x and g refer to all values in continuous column C, and X and G, represent the distribution for real and generated data, respectively. In this score, the correlation between the columns is bounded between -1 and 1, with -1 representing the most negative correlation and 1 representing the most positive correlation between the real and generated column.\\
Traditionally, relational databases are divided into separate tables for each object, and to retrieve information from those tables, related fields within the tables need to be linked together by defining relationships. Users can then retrieve information from multiple tables simultaneously by calling queries \cite{date2019database}. In terms of measuring the similarity among those tables, the cardinality of related tables can be used. The cardinality of a table relationship is determined by how many rows in each table are related. Therefore, when generating synopses, measuring whether a table's cardinality is the same between the real and generated data is an important metric. In order to measure the similarity of cardinality between related tables, marginal distribution can be utilized by computing the similarity between a real and generated cardinality distribution using KS or TVD score. In the case of real and generated data, the cardinality complement score returns 1 when the cardinality values are the same and 0 when they are different.

%% file: sections/10-Conclusion.tex
The construction of synopses is essential to data-driven decision-making systems in order to provide approximate answers to queries. Since traditional statistical approaches are ineffective, many researchers are exploring how realistic data can be generated with Deep Generative Models in order to revolutionize the AQP system. In this paper, we discussed the challenges associated with the generation of synopses in relational databases and whether Generative Adversarial Networks can be used to accomplish this task. Furthermore, we summarized and reformed statistical metrics for evaluating the quality of the generated synopses as a part of this study.\\
There is no doubt that GANs have an incredible ability to generate realistic images and videos. However, each data point in an image is represented by a pixel, which cannot be interpreted alone but only in relation to the other pixels in the image. Consequently, the meaning of the same pixel in one image differs from the meaning of the corresponding pixel in another image \cite{lederrey2022datgan}. In contrast with images, data tables typically contain columns that have a specific meaning and can be understood by their positions and values, and their values may also have a semantic relationship with each other. We analyzed the challenges associated with synopsis construction in a relational database and categorized them into the following categories: data type, bounded continuous columns, non-Gaussian distribution, imbalance categorical columns, and semantic relationship and constraint between columns. Then, by reviewing the promising variants of GANs designed for generating tabular data, we realized that the solutions to the given challenges revolve around the following areas. First, data transformation in preprocessing phase, especially for handling categorical and null values. Second, data distribution matching, typically by defining specific loss functions to penalize discriminator and generator for the difference between generated and real data distribution to learn multimodal mapping from inputs to outputs. Third, conditional and informed generator where the generator is conditioned upon some sort of auxiliary information (such as class labels or data) from other modalities so that the generator is fed with different contextual information, and prior knowledge, so that it can capture the interactions between columns in a data set. We demonstrated that although the majority of proposed methods are geared towards applications such as data privacy for data sharing, data augmentation for machine learning model training, and data imputation for missing values, rather than generating synopsis, GANs are capable of generating data synopsis that are identical to actual data.\\
In summary, Generative Adversarial Networks have demonstrated tremendous potential in the world of machine learning for creating realistic images, videos, audio, and text. However, the complexity of tabular data makes it difficult for algorithms to understand semantic relationships and constraints in relational databases during the training process. The field of GANs and Adversarial Learning is still relatively young, thus, there is a need to improve current methods in order to be able to construct effective synopses for AQP in data-driven decision-making systems.

%% file: references.bib
@inproceedings{sagiroglu2013big,
  title={Big data: A review},
  author={Sagiroglu, Seref and Sinanc, Duygu},
  booktitle={2013 international conference on collaboration technologies and systems (CTS)},
  pages={42--47},
  year={2013},
  organization={IEEE}
}

@article{li2018approximate,
  title={Approximate query processing: What is new and where to go?},
  author={Li, Kaiyu and Li, Guoliang},
  journal={Data Science and Engineering},
  volume={3},
  number={4},
  pages={379--397},
  year={2018},
  publisher={Springer}
}

@inproceedings{muniswamaiah2020approximate,
  title={Approximate Query Processing for Big Data in Heterogeneous Databases},
  author={Muniswamaiah, Manoj and Agerwala, Tilak and Tappert, Charles C},
  booktitle={2020 IEEE International Conference on Big Data (Big Data)},
  pages={5765--5767},
  year={2020},
  organization={IEEE}
}

@inproceedings{chaudhuri2017approximate,
  title={Approximate query processing: No silver bullet},
  author={Chaudhuri, Surajit and Ding, Bolin and Kandula, Srikanth},
  booktitle={Proceedings of the 2017 ACM International Conference on Management of Data},
  pages={511--519},
  year={2017}
}

@misc{liu2009approximate,
  title={Approximate Query Processing.},
  author={Liu, Qing},
  year={2009}
}

@inproceedings{ma2019dbest,
  title={Dbest: Revisiting approximate query processing engines with machine learning models},
  author={Ma, Qingzhi and Triantafillou, Peter},
  booktitle={Proceedings of the 2019 International Conference on Management of Data},
  pages={1553--1570},
  year={2019}
}

@article{zhang2021laqp,
  title={LAQP: Learning-based approximate query processing},
  author={Zhang, Meifan and Wang, Hongzhi},
  journal={Information Sciences},
  volume={546},
  pages={1113--1134},
  year={2021},
  publisher={Elsevier}
}

@article{savva2020ml,
  title={Ml-aqp: Query-driven approximate query processing based on machine learning},
  author={Savva, Fotis and Anagnostopoulos, Christos and Triantafillou, Peter},
  journal={arXiv preprint arXiv:2003.06613},
  year={2020}
}

@article{ruthotto2021introduction,
  title={An introduction to deep generative modeling},
  author={Ruthotto, Lars and Haber, Eldad},
  journal={GAMM-Mitteilungen},
  volume={44},
  number={2},
  pages={e202100008},
  year={2021},
  publisher={Wiley Online Library}
}

@inproceedings{thirumuruganathan2020approximate,
  title={Approximate query processing for data exploration using deep generative models},
  author={Thirumuruganathan, Saravanan and Hasan, Shohedul and Koudas, Nick and Das, Gautam},
  booktitle={2020 IEEE 36th international conference on data engineering (ICDE)},
  pages={1309--1320},
  year={2020},
  organization={IEEE}
}

@Inbook{Markl2009,
author="Markl, Volker",
title="Query Processing (in Relational Databases)",
bookTitle="Encyclopedia of Database Systems",
year="2009",
publisher="Springer US",
address="Boston, MA",
pages="2288--2293",
isbn="978-0-387-39940-9",
doi="10.1007/978-0-387-39940-9\_296"
}

@inproceedings{hellerstein1997online,
  title={Online aggregation},
  author={Hellerstein, Joseph M and Haas, Peter J and Wang, Helen J},
  booktitle={Proceedings of the 1997 ACM SIGMOD international conference on Management of data},
  pages={171--182},
  year={1997}
}

@article{mozafari2015handbook,
  title={A Handbook for Building an Approximate Query Engine.},
  author={Mozafari, Barzan and Niu, Ning},
  journal={IEEE Data Eng. Bull.},
  volume={38},
  number={3},
  pages={3--29},
  year={2015}
}

@Inbook{Zhang2009,
author="Zhang, Qing",
title="Data Sampling",
bookTitle="Encyclopedia of Database Systems",
year="2009",
publisher="Springer US",
address="Boston, MA",
pages="630--634",
isbn="978-0-387-39940-9",
doi="10.1007/978-0-387-39940-9\_535"
}

@article{piatetsky1984accurate,
  title={Accurate estimation of the number of tuples satisfying a condition},
  author={Piatetsky-Shapiro, Gregory and Connell, Charles},
  journal={ACM Sigmod Record},
  volume={14},
  number={2},
  pages={256--276},
  year={1984},
  publisher={ACM New York, NY, USA}
}

@article{RUSSELL20082316,
title = {Applications of wavelet data reduction in a recommender system},
journal = {Expert Systems with Applications},
volume = {34},
number = {4},
pages = {2316-2325},
year = {2008},
issn = {0957-4174},
doi = {https://doi.org/10.1016/j.eswa.2007.03.009},
author = {Stephen Russell and Victoria Yoon},
}

@inproceedings{dell2010accuracy,
  title={Accuracy estimation in approximate query processing},
  author={Dell'Aquila, Carlo and Di Tria, Francesco and Lefons, Ezio and Tangorra, Filippo},
  booktitle={Proceedings of the 14th WSEAS international conference on Computers: part of the 14th WSEAS CSCC multiconference-Volume II},
  pages={452--458},
  year={2010}
}

@inproceedings{yang2017sf,
  title={Sf-sketch: A fast, accurate, and memory efficient data structure to store frequencies of data items},
  author={Yang, Tong and Liu, Lingtong and Yan, Yibo and Shahzad, Muhammad and Shen, Yulong and Li, Xiaoming and Cui, Bin and Xie, Gaogang},
  booktitle={2017 IEEE 33rd International Conference on Data Engineering (ICDE)},
  pages={103--106},
  year={2017},
  organization={IEEE}
}

@article{halevy2001answering,
  title={Answering queries using views: A survey},
  author={Halevy, Alon Y},
  journal={The VLDB Journal},
  volume={10},
  number={4},
  pages={270--294},
  year={2001},
  publisher={Springer}
}

@inproceedings{spiegel2006graph,
  title={Graph-based synopses for relational selectivity estimation},
  author={Spiegel, Joshua and Polyzotis, Neoklis},
  booktitle={Proceedings of the 2006 ACM SIGMOD international conference on Management of data},
  pages={205--216},
  year={2006}
}

@article{spiegel2009tug,
  title={TuG synopses for approximate query answering},
  author={Spiegel, Joshua and Polyzotis, Neoklis},
  journal={ACM Transactions on Database Systems (TODS)},
  volume={34},
  number={1},
  pages={1--56},
  year={2009},
  publisher={ACM New York, NY, USA}
}

@incollection{aggarwal2007survey,
  title={A survey of synopsis construction in data streams},
  author={Aggarwal, Charu C and Yu, Philip S},
  booktitle={Data streams},
  pages={169--207},
  year={2007},
  publisher={Springer}
}

@article{tan2022one,
  title={One-pass streaming algorithm for DR-submodular maximization with a knapsack constraint over the integer lattice},
  author={Tan, Jingjing and Zhang, Dongmei and Zhang, Hongyang and Zhang, Zhenning},
  journal={Computers and Electrical Engineering},
  volume={99},
  pages={107766},
  year={2022},
  publisher={Elsevier}
}

@article{goodfellow2016nips,
  title={Nips 2016 tutorial: Generative adversarial networks},
  author={Goodfellow, Ian},
  journal={arXiv preprint arXiv:1701.00160},
  year={2016}
}

@article{gui2021review,
  title={A review on generative adversarial networks: Algorithms, theory, and applications},
  author={Gui, Jie and Sun, Zhenan and Wen, Yonggang and Tao, Dacheng and Ye, Jieping},
  journal={IEEE Transactions on Knowledge and Data Engineering},
  year={2021},
  publisher={IEEE}
}

@article{goodfellow2014generative,
  title={Generative adversarial nets},
  author={Goodfellow, Ian and Pouget-Abadie, Jean and Mirza, Mehdi and Xu, Bing and Warde-Farley, David and Ozair, Sherjil and Courville, Aaron and Bengio, Yoshua},
  journal={Advances in neural information processing systems},
  volume={27},
  year={2014}
}

@article{xu2018synthesizing,
  title={Synthesizing tabular data using generative adversarial networks},
  author={Xu, Lei and Veeramachaneni, Kalyan},
  journal={arXiv preprint arXiv:1811.11264},
  year={2018}
}

@article{xu2019modeling,
  title={Modeling tabular data using conditional gan},
  author={Xu, Lei and Skoularidou, Maria and Cuesta-Infante, Alfredo and Veeramachaneni, Kalyan},
  journal={Advances in Neural Information Processing Systems},
  volume={32},
  year={2019}
}

@article{khurana2020semantic,
  title={Semantic Annotation for Tabular Data},
  author={Khurana, Udayan and Galhotra, Sainyam},
  journal={arXiv preprint arXiv:2012.08594},
  year={2020}
}

@article{lederrey2022datgan,
  title={DATGAN: Integrating expert knowledge into deep learning for synthetic tabular data},
  author={Lederrey, Gael and Hillel, Tim and Bierlaire, Michel},
  journal={arXiv preprint arXiv:2203.03489},
  year={2022}
}

@inproceedings{arjovsky2017wasserstein,
  title={Wasserstein generative adversarial networks},
  author={Arjovsky, Martin and Chintala, Soumith and Bottou, L{\'e}on},
  booktitle={International conference on machine learning},
  pages={214--223},
  year={2017},
  organization={PMLR}
}

@article{gulrajani2017improved,
  title={Improved training of wasserstein gans},
  author={Gulrajani, Ishaan and Ahmed, Faruk and Arjovsky, Martin and Dumoulin, Vincent and Courville, Aaron C},
  journal={Advances in neural information processing systems},
  volume={30},
  year={2017}
}

@inproceedings{zhao2021ctab,
  title={Ctab-gan: Effective table data synthesizing},
  author={Zhao, Zilong and Kunar, Aditya and Birke, Robert and Chen, Lydia Y},
  booktitle={Asian Conference on Machine Learning},
  pages={97--112},
  year={2021},
  organization={PMLR}
}

@article{park2018data,
  title={Data synthesis based on generative adversarial networks},
  author={Park, Noseong and Mohammadi, Mahmoud and Gorde, Kshitij and Jajodia, Sushil and Park, Hongkyu and Kim, Youngmin},
  journal={arXiv preprint arXiv:1806.03384},
  year={2018}
}

@inproceedings{odena2017conditional,
  title={Conditional image synthesis with auxiliary classifier gans},
  author={Odena, Augustus and Olah, Christopher and Shlens, Jonathon},
  booktitle={International conference on machine learning},
  pages={2642--2651},
  year={2017},
  organization={PMLR}
}

@manual{
    sdmetrics,
    title = {Synthetic Data Metrics},
    organization = {DataCebo, Inc.},
    year = {2022},
    month = {9},
    note = {v0.7.0},
    url = {https://docs.sdv.dev/sdmetrics/}
}

@book{date2019database,
  title={Database design and relational theory: normal forms and all that jazz},
  author={Date, Chris J},
  year={2019},
  publisher={Apress}
}

@incollection{biskup1981formal,
  title={A formal approach to null values in database relations},
  author={Biskup, Joachim},
  booktitle={Advances in Data Base Theory},
  pages={299--341},
  year={1981},
  publisher={Springer}
}

@article{wang2021generative,
  title={Generative adversarial networks in computer vision: A survey and taxonomy},
  author={Wang, Zhengwei and She, Qi and Ward, Tomas E},
  journal={ACM Computing Surveys (CSUR)},
  volume={54},
  number={2},
  pages={1--38},
  year={2021},
  publisher={ACM New York, NY, USA}
}

@inproceedings{choi2017generating,
  title={Generating multi-label discrete patient records using generative adversarial networks},
  author={Choi, Edward and Biswal, Siddharth and Malin, Bradley and Duke, Jon and Stewart, Walter F and Sun, Jimeng},
  booktitle={Machine learning for healthcare conference},
  pages={286--305},
  year={2017},
  organization={PMLR}
}

@article{mottini2018airline,
  title={Airline passenger name record generation using generative adversarial networks},
  author={Mottini, Alejandro and Lheritier, Alix and Acuna-Agost, Rodrigo},
  journal={arXiv preprint arXiv:1807.06657},
  year={2018}
}

@article{bellemare2017cramer,
  title={The cramer distance as a solution to biased wasserstein gradients},
  author={Bellemare, Marc G and Danihelka, Ivo and Dabney, Will and Mohamed, Shakir and Lakshminarayanan, Balaji and Hoyer, Stephan and Munos, R{\'e}mi},
  journal={arXiv preprint arXiv:1705.10743},
  year={2017}
}

@article{radford2015unsupervised,
  title={Unsupervised representation learning with deep convolutional generative adversarial networks},
  author={Radford, Alec and Metz, Luke and Chintala, Soumith},
  journal={arXiv preprint arXiv:1511.06434},
  year={2015}
}

@article{mirza2014conditional,
  title={Conditional generative adversarial nets},
  author={Mirza, Mehdi and Osindero, Simon},
  journal={arXiv preprint arXiv:1411.1784},
  year={2014}
}

@article{theis2015note,
  title={A note on the evaluation of generative models},
  author={Theis, Lucas and Oord, A{\"a}ron van den and Bethge, Matthias},
  journal={arXiv preprint arXiv:1511.01844},
  year={2015}
}

@article{gretton2006kernel,
  title={A kernel method for the two-sample-problem},
  author={Gretton, Arthur and Borgwardt, Karsten and Rasch, Malte and Sch{\"o}lkopf, Bernhard and Smola, Alex},
  journal={Advances in neural information processing systems},
  volume={19},
  year={2006}
}

@article{deecke2018mode,
  title={Mode normalization},
  author={Deecke, Lucas and Murray, Iain and Bilen, Hakan},
  journal={arXiv preprint arXiv:1810.05466},
  year={2018}
}
